\documentclass[preprint2]{emulateapj} 
\usepackage{amsmath}
\usepackage{natbib}
\usepackage{graphicx}
\usepackage{epsf}
\usepackage{color}
\input{epsf}
\usepackage{epsfig}
\DeclareGraphicsExtensions{.jpg,.pdf,.png,.eps,.ps}
\graphicspath{{FIGURES/}}

\newcommand{\ltsima}{$\; \buildrel < \over \sim \;$}
\newcommand{\ltsim}{\lower.5ex\hbox{\ltsima}}

\def\plottwo#1#2{\centering \leavevmode
   \epsfxsize=1.0\columnwidth \epsfbox{#1} \hfil
   \epsfxsize=1.0\columnwidth \epsfbox{#2}}

\begin{document}

\title{Measurements of Secondary Cosmic Microwave Background Anisotropies with the South Pole Telescope}

\author{
M.~Lueker,\altaffilmark{1}
C.~L.~Reichardt,\altaffilmark{1}
K.~K.~Schaffer,\altaffilmark{2,3} 
O.~Zahn,\altaffilmark{4} 
P.~A.~R.~Ade,\altaffilmark{5}
K.~A.~Aird,\altaffilmark{6}
B.~A.~Benson,\altaffilmark{1,2,3}
L.~E.~Bleem,\altaffilmark{2,7} 
J.~E.~Carlstrom,\altaffilmark{2,3,7,8} 
C.~L.~Chang,\altaffilmark{2,3} 
H.-M. Cho,\altaffilmark{1}  
T.~M.~Crawford,\altaffilmark{2,8} 
A.~T.~Crites,\altaffilmark{2,8} 
T.~de~Haan,\altaffilmark{9} 
M.~A.~Dobbs,\altaffilmark{9}
E.~M.~George,\altaffilmark{1}
N.~R.~Hall,\altaffilmark{10}
N.~W.~Halverson,\altaffilmark{11} 
G.~P.~Holder,\altaffilmark{9}
W.~L.~Holzapfel,\altaffilmark{1} 
J.~D.~Hrubes,\altaffilmark{6}
M.~Joy,\altaffilmark{12}
R.~Keisler,\altaffilmark{2,7} 
L.~Knox,\altaffilmark{10}
A.~T.~Lee,\altaffilmark{1,13} 
E.~M.~Leitch,\altaffilmark{2,8}
J.~J.~McMahon,\altaffilmark{2,3,14} 
J.~Mehl,\altaffilmark{1}
S.~S.~Meyer,\altaffilmark{2,3,7,8}
J.~J.~Mohr,\altaffilmark{15, 16, 17}
T.~E.~Montroy,\altaffilmark{18} 
S.~Padin,\altaffilmark{2,8}
T.~Plagge,\altaffilmark{1}
C.~Pryke,\altaffilmark{2,3,8} 
J.~E.~Ruhl,\altaffilmark{18} 
L.~Shaw,\altaffilmark{9,19}
 E.~Shirokoff,\altaffilmark{1} 
H.~G.~Spieler,\altaffilmark{13}
Z.~Staniszewski,\altaffilmark{18} 
A.~A.~Stark,\altaffilmark{20} 
K.~Vanderlinde,\altaffilmark{9} 
J.~D.~Vieira,\altaffilmark{2,7} and
R.~Williamson\altaffilmark{2,8} 
}

\altaffiltext{1}{Department of Physics,
University of California, Berkeley, CA 94720}
\altaffiltext{2}{Kavli Institute for Cosmological Physics,
University of Chicago,
5640 South Ellis Avenue, Chicago, IL 60637}
\altaffiltext{3}{Enrico Fermi Institute,
University of Chicago,
5640 South Ellis Avenue, Chicago, IL 60637}
\altaffiltext{4}{Berkeley Center for Cosmological Physics,
Department of Physics, University of California, and Lawrence Berkeley
National Labs, Berkeley, CA 94720}
\altaffiltext{5}{Department of Physics and Astronomy,
Cardiff University, CF24 3YB, UK}
\altaffiltext{6}{University of Chicago,
5640 South Ellis Avenue, Chicago, IL 60637}
\altaffiltext{7}{Department of Physics,
University of Chicago,
5640 South Ellis Avenue, Chicago, IL 60637}
\altaffiltext{8}{Department of Astronomy and Astrophysics,
University of Chicago,
5640 South Ellis Avenue, Chicago, IL 60637}
\altaffiltext{9}{Department of Physics,
McGill University,
3600 Rue University, Montreal, Quebec H3A 2T8, Canada}
\altaffiltext{10}{Department of Physics, 
University of California, One Shields Avenue, Davis, CA 95616}
\altaffiltext{11}{Department of Astrophysical and Planetary Sciences and Department of Physics,
University of Colorado,
Boulder, CO 80309}
\altaffiltext{12}{Department of Space Science, VP62,
NASA Marshall Space Flight Center,
Huntsville, AL 35812}
\altaffiltext{13}{Physics Division,
Lawrence Berkeley National Laboratory,
Berkeley, CA 94720}
\altaffiltext{14}{Department of Physics, University of Michigan, 450 Church Street, Ann  
Arbor, MI, 48109}
\altaffiltext{15}{Department of Physics,
Ludwig-Maximilians-Universit\"{a}t,
Scheinerstr.\ 1, 81679 M\"{u}nchen, Germany}
\altaffiltext{16}{Excellence Cluster Universe,
Boltzmannstr.\ 2, 85748 Garching, Germany}
\altaffiltext{17}{Max-Planck-Institut f\"{u}r extraterrestrische Physik,
Giessenbachstr.\ 85748 Garching, Germany}
\altaffiltext{18}{Physics Department, Center for Education and Research in Cosmology 
and Astrophysics, 
Case Western Reserve University,
Cleveland, OH 44106}
\altaffiltext{19}{Department of Physics, Yale University, P.O. Box 208210, New Haven,
CT 06520-8120}
\altaffiltext{20}{Harvard-Smithsonian Center for Astrophysics,
60 Garden Street, Cambridge, MA 02138}

\email{lueker@socrates.berkeley.edu}
 
\begin{abstract}
We report 
cosmic microwave background (CMB) power spectrum
measurements from the first 100~deg$^2$ field observed by the South Pole Telescope (SPT)  at 150 and 
220$\,$GHz. On angular scales where the primary CMB anisotropy is dominant, 
$\ell \lesssim 3000$,  the SPT power spectrum is consistent with the standard 
$\Lambda$CDM cosmology. On smaller scales, we see strong evidence for
 a point source contribution, consistent with a population of dusty, star-forming galaxies. 
After we mask bright point sources, anisotropy power on angular scales 
of $3000 < \ell < 9500$ is detected with a signal-to-noise $\gtrsim 50$ 
at both frequencies.  
We combine the 150 and 
220$\,$GHz data to remove the majority of the point source power, and use the 
point source subtracted spectrum to detect Sunyaev-Zel'dovich (SZ) power at $2.6\,\sigma$.
At $\ell=3000$, the SZ power in the subtracted bandpowers is $4.2\,$$\pm$$\,1.5\, \mu\rm{K}^2$, which is significantly lower than the power predicted by a fiducial model using WMAP5 cosmological parameters. 
This discrepancy may suggest that contemporary galaxy cluster models overestimate the thermal pressure of intracluster gas.
Alternatively, this result can be interpreted as evidence for lower values of $\sigma_8$.
When combined with an estimate of the kinetic SZ contribution,
the measured SZ amplitude shifts $\sigma_8$ from the primary CMB anisotropy derived constraint of $0.794\,$$\pm$$\,0.028$ down to $0.773\,$$\pm$$\,0.025$. The uncertainty in the constraint on $\sigma_8$ from this analysis is dominated by uncertainties in the theoretical modeling required to predict the amplitude of the SZ power spectrum for a given set of cosmological parameters.

\end{abstract}

\keywords{cosmology -- cosmology:cosmic microwave background --  cosmology: observations -- large-scale structure of universe }

\bigskip\bigskip

\section{Introduction}
\label{sec:intro}

Anisotropy in the cosmic microwave background (CMB) has been
well-characterized on angular scales larger than a few
arcminutes \nolinebreak{\citep{jones06,reichardt09a,nolta09,brown09}}, providing
stringent tests of cosmological theory and strong evidence for the
$\Lambda$CDM cosmological model \citep{hinshaw09,komatsu09}. 
The primary CMB anisotropy is a direct record
of density fluctuations at the surface of last scattering.  On angular scales $\lesssim 10$ arcminutes, the
primary CMB anisotropy is exponentially damped due to photon diffusion in
the primordial plasma \citep{silk68}; the resulting decline in power with increasing multipole is known as the ``damping tail". 
The anisotropy
on very small scales, which
is only beginning to be explored experimentally, is
instead dominated by foreground emission and secondary distortions to
the CMB introduced after the surface of last scattering.  This
secondary CMB anisotropy is produced by interactions of CMB photons
with large scale structure and can 
potentially provide sensitive and independent tests of cosmological
parameters important to structure formation.   

The most significant source of secondary CMB temperature anisotropy on small
angular scales is predicted to be the inverse Compton scattering of CMB photons by 
the hot plasma gravitationally bound to 
massive, collapsed objects \citep{sunyaev72}.  This is known as the thermal
Sunyaev-Zel'dovich (tSZ) effect.  
The tSZ effect distorts the blackbody spectrum of the primary CMB, creating a decrement
in intensity at low frequencies and an increment at high frequencies with a null near 220$\,$GHz.
At 150$\,$GHz, the
anisotropy induced by the tSZ effect is expected to dominate over the
primary CMB fluctuations at multipoles
$\ell\gtrsim3000$. 
The power contributed to CMB maps by the tSZ effect depends sensitively on the normalization
of the matter power spectrum, as parametrized by the RMS of the mass distribution on 8h$^{-1}$ Mpc
scales, $\sigma_8$. 

Galaxy clusters over a wide range of mass and redshift
contribute significantly to the tSZ power spectrum.
It is challenging to accurately model the low-mass or high-redshift 
clusters.
For instance, lower mass clusters
are more sensitive to non-gravitational heating effects such as by AGN,
while we have limited observational data on high-redshift clusters.
There is considerable uncertainty as to the expected shape and amplitude
of the tSZ power spectrum due to the current
lack of knowledge of the properties of intracluster gas in low mass and
high redshift galaxy
clusters.

The scattered CMB photons also obtain a net Doppler shift when ionized
matter is moving with respect to the rest frame of the CMB.
This
so-called kinetic Sunyaev-Zel'dovich (kSZ) effect \citep{sunyaev80b}
depends solely on the motion and density of free electrons. 
In contrast to the tSZ effect, the kSZ effect has contributions from electrons with temperatures as low as 10$^4\,$K.
Therefore higher-redshift epochs, before massive objects finish collapsing, are expected to have relatively larger 
contributions to the kSZ power.
Recent simulations and analytic models also predict a sizable signal from
the epoch of the first radiative sources which form ionized regions several tens of Mpc across, 
within a largely neutral Universe. 
Low-redshift galaxy clusters dominate the power on small angular scales, while
high-redshift reionizing regions have their
largest relative contribution on angular scales around $\ell = 2000$.
At 150$\,$GHz, the kinetic effect is expected to amount to tens of percent
of the total SZ power.

Only a handful of experiments have had sufficient sensitivity and angular resolution
to probe the damping tail of the CMB anisotropy.   
Early measurements at 30$\,$GHz by CBI \citep{mason03,bond05} 
reported a $>\,$$3$$\,\sigma$ excess above the expected CMB
power at multipoles of $\ell > 2000$.  
Observations with the BIMA array at $30\,$GHz \citep{dawson06} 
also reported a nearly $2$$\,\sigma$ detection of excess power at $\ell=5237$.  
However, more recently, the SZA
experiment (also observing at 30$\,$GHz) has published an upper limit of $149\,\mu$K$^2$ at 
95\% confidence on excess power at these multipoles \citep{sharp09} in apparent conflict with 
the previous CBI and BIMA results. 
For the relatively small patch ($0.1\,{\rm deg}^2$) observed by BIMA, the non-Gaussian nature 
of the SZ sky means that there is no significant tension between the BIMA and SZA results. 
The latest CBI measurements \citep{sievers09} include more data, improved
radio source removal, and a proper treatment of non-Gaussianity of the SZ sky. 
These measurements continue to suggest excess power but with a significance of only $1.6\,\sigma$.

At 150$\,$GHz, the ACBAR \citep{reichardt09a} and QUaD \citep{friedman09}
experiments have both measured the damping tail of the primary CMB
anisotropy at $\ell < 3000$ with high signal to noise. 
Either with or without the addition of the expected foreground and tSZ contributions, the power measured at the
highest multipoles by both experiments is consistent with primary CMB anisotropy. 
In the last year, the results of 150$\,$GHz observations out to $\ell = 10000$ made with the Bolocam
\citep{sayers09} and APEX-SZ \citep{reichardt09b} experiments have been released. 
These experiments have been used to place upper limits on power above the primary CMB of $1080\,\mu$K$^2$ 
and $105\,\mu$K$^2$ respectively at 95\% confidence. 
The constraints on $\sigma_8$ from these upper limits remain weak, in no small part due to the large, 
highly non-Gaussian sample variance of the tSZ effect on the small $\sim$1 deg$^2$ patches of sky 
observed by Bolocam and APEX-SZ. 
The cosmic variance of the tSZ effect will be significantly reduced in the on-going $\gtrsim\,100\,$ deg$^2$ surveys being conducted by next-generation experiments such as ACT \citep{fowler07} and SPT.

In addition to the tSZ and kSZ effects, foreground emission is important on these small angular scales. 
After bright radio sources are removed, the most significant foreground at 150 and 220$\,$GHz is expected to be a population
of unresolved, faint,
dusty, star-forming galaxies (DSFGs) with a rest frame emission spectrum that peaks in
the far infrared.  
These sources have been studied at higher frequencies close to the peak of their emission spectrum\footnote{Since these sources are typically brightest at sub-millimeter wavelengths they are also referred to in the literature as sub-millimeter galaxies (SMGs).} (e.g. \cite{holland99,kreysa98,glenn98,viero09}), however extrapolating their fluxes to 150$\,$GHz remains uncertain.  
Adding to the challenge is the expected significant clustering of these sources
\citep{haiman00,knox01,righi08,sehgal09}. 
IR emission from clustered DSFGs was first observed with the Spitzer telescope at 160$\,\mu$m \citep{lagache07} and more recently this clustering has also been observed at sub-mm wavelengths by the BLAST experiment 
\citep{viero09}. 
The clustering of these DSFGs is expected to produce anisotropic power at 150$\,$GHz with an angular power spectrum that
is similar to that of the SZ effect.  However, emission from DSFGs is spectrally separable from the 
SZ effect and the SZ power spectrum can be recovered by combining information from 
overlapping maps at 150 and 220$\,$GHz.  

In this work, we present measurements by the South Pole Telescope
(SPT) which comprise the first significant detections of anisotropy power for $\ell > 3000$ at 150 and 220$\,$GHz.
The SPT has sufficient angular resolution, sensitivity and sky coverage to 
produce high-precision measurements of anisotropy over a range of multipoles from 
$\sim 100 < \ell < 9500$.
However, for the immediate goal of measuring secondary CMB anisotropies, we start with the first bandpower 
at $\ell=2000$ where primary CMB still dominates the power spectrum.
We combine bandpowers 
from two frequencies to minimize the DSFG contribution
and produce the  first significant detection of the SZ contribution to the CMB
power spectrum.  In a companion paper \citet[][hereafter H09]{hall09}, the SPT power spectra are used
to place constraints on the amplitude of the Poisson and
clustered components of DSFG power. 
This is the first detection of the clustered DSFG power at mm-wavelengths.
We use three results from H09 in setting limits on the residual point source power in the DSFG-subtracted bandpowers (\S\ref{sec:ptsrcsub}): an argument that residual clustered components are negligible, 
a measurement of the Poisson point source power at 150$\,$GHz, and a limit on the
scatter in spectral index between DSFGs.  

We describe the instrument, observations, beams, and calibration strategy in \S\ref{sec:obs-reduc}. 
The time-ordered data (TOD) filtering and map-making algorithm is outlined in \S\ref{sec:analysis}, along with the procedure to derive bandpowers from maps. 
The results of tests for systematic errors applied to the SPT data are discussed in \S\ref{sec:jackknives}, and the expected astronomical foregrounds are described in \S\ref{sec:foregrounds}. 
The bandpowers and cosmological interpretation are given in \S\ref{sec:results}.

\section{Instrument and Observations}
\label{sec:obs-reduc}

The SPT is an off-axis Gregorian telescope with a 10-m diameter
primary mirror located at the South Pole.  The telescope is optimized to perform high resolution 
surveys of low surface brightness sources.
The first receiver to be placed on the SPT has the primary goal of identifying a mass-limited sample of galaxy clusters, the first of which were reported 
in \citet[][hereafter S09]{staniszewski08}.  
The receiver is equipped with a 960-element array of 
superconducting transition edge sensor bolometers.
The detectors are split between three frequency bands centered at
$95\,$GHz, $150\,$GHz, and $220\,$GHz, allowing the separation of the tSZ effect
from the primary CMB anisotropy and foregrounds. Further details of
the receiver and telescope can be found in \citet{ruhl04},
\citet{padin08} and \citet{carlstrom09}.

In this work, we use data at 150 and 220$\,$GHz from one 100 deg$^2$ field
centered at right ascension $5^\mathrm{h} 30^\mathrm{m}$, 
declination $-55^\circ$ (J2000) observed by SPT in the first half of
the 2008 austral winter. 
Data from SPT's third (95 GHz) frequency band was unusable in 2008. 
The location of the field was chosen for 
overlap with the Blanco Cosmology Survey (BCS)\footnote{http://cosmology.illinois.edu/BCS} optical survey and low dust emission.  We observed this field for a total of 779 hours, of which 575
hours is used in the analysis after passing data quality cuts. 
 The
final map noise is 18$\,\mu\textrm{K}$-arcmin\footnote{Throughout this work, the unit $\textrm{K}$ refers to equivalent fluctuations in the CMB temperature, i.e.,~the level of temperature fluctuation of a 2.73$\,$K blackbody that would be required to produce the same power fluctuation.  
The conversion factor is given by the derivative of the blackbody spectrum, $\frac{dB}{dT}$, evaluated at 2.73$\,$K.} 
at 150$\,$GHz and 40$\,\mu\textrm{K}$-arcmin at 220$\,$GHz. 
This field accounts for half the sky area observed in 2008 and one eighth of the total area observed by SPT to date.

The scanning strategy used for these observations involves
constant-elevation scans across the 10$^\circ$ wide field. After each scan back
and forth across the field, the telescope executes a 0.125$^\circ$ step
in elevation.
A complete set of scans covering the entire field takes approximately two hours, and we
refer to each complete set as an observation.  Successive
observations use a
series of different starting elevations to ensure even coverage of the field.
Of the 300 observations used in this analysis,  half 
were performed at an azimuth scanning speed of 0.44$^\circ$ per second and half at a speed of 0.48$^\circ$ per second.

\subsection{Beam Measurements}
\label{sec:beams}

The SPT beams are measured by combining maps of three sources: 
Jupiter, Venus, and the brightest point source in the 100 deg$^2$ field.
Observations of Jupiter 
are used to measure a diffuse, low-level sidelobe  
in the range $15^\prime<\rm{r}<40^\prime$, where r is the radius to the beam center.  
Although this sidelobe has a low amplitude ($-50\,$dB at r=$30^\prime$), 
it contains approximately 15\% of the total beam solid angle.  
A measurement of this sidelobe is necessary for the 
cross-calibration with WMAP described in \S \ref{sec:calibration}.  
The observations of Jupiter show signs of potential non-linearity in
the response of the detectors for r $<10^\prime$.  For this reason we
only use the observations of Jupiter to map the sidelobe at r $>15^\prime$.
Observations of Venus are used to measure the beam in 
the region 4$^\prime<$ r $<15^\prime$.  
The angular extent of Jupiter or Venus has a negligible effect on the 
measurement of the relatively smooth beam features present at these large radii.
The brightest point source in the map of the 100 deg$^2$ field is used to 
measure the beam within a radius of 4$^\prime$.  
In this way, the random error in the pointing reconstruction 
($7^{\prime\prime}$ RMS) and its impact on the effective beam are taken into account.  
The pointing error has a negligible effect on the relatively smooth outer 
(r $>4^\prime$) region of the beam.

A composite beam map, $B(\theta,\phi)$, is assembled by merging maps of Jupiter, Venus,
and the bright point source. The TOD 
are filtered prior to making these maps, in order  
to remove large-scale atmospheric noise.  
Masks with radii of 40$^\prime$, 25$^\prime$,  and 5$^\prime$ are applied around the locations of Jupiter, Venus, and the point source, respectively. 
These masks ensure that the beam measurements are not affected by the
filtering.

Using the the flat-sky approximation, we calculate the Fourier transform of
the composite beam map, $B(\ell,\phi_\ell)$.
From this, we compute the azimuthally-averaged beam function,
\begin{equation}
  B_\ell = \rm{Re}\{\frac{1}{2\pi}\int B(\ell,\phi_\ell)d\phi_\ell\}.
\end{equation}
We note that averaging $|B(\ell,\phi_\ell)|^2$ instead of $B(\ell,\phi_\ell)$ would result in a percent-level noise bias in $B_\ell$ at very high multipoles due to the presence of noise.
The results in this work assume an axially symmetric beam, 
which is only an approximation for SPT.  
We simulate the effects of ignoring the asymmetry on the bandpowers and find that the errors introduced by making this 
assumption are negligible.

Although the measured beam function $B_\ell$ is used for the bandpower estimation, 
an empirical fit is used to quantify the errors on $B_\ell$.
$B_\ell$ is fit to the empirical model
\begin{equation}
  B_\ell = ae^{-\frac{1}{2}(\sigma_b\ell)^{1.5}} + (1-a)e^{-\frac{1}{2}(0.00292*\ell)^{1.8}}.
\end{equation}
There are two components: a main lobe (first term) and a diffuse shelf (second term).  
The form of the model and the numerical values of the slopes of the exponents were 
constructed to provide a good fit to the measured $B_\ell$.  
We note that $B_{\ell}^{\rm{150}}$ and $B_{\ell}^{\rm{220}}$ are measured and fitted separately.
The RMS difference between the model and measured $B_\ell$s is approximately 1\%.
Two parameters remain free: 
$\sigma_b$, which describes the width of the main lobe, and 
$a$, which sets the relative normalization between the main lobe and the diffuse shelf.  
These parameters are left free to quantify the uncertainty in $B_\ell$.  
The uncertainty in the values of these parameters directly translates to an uncertainty 
in $B_\ell$.  

There are a number of factors that limit the accuracy of the measurement of $B_\ell$.  
These include residual map noise, errors associated with the map-merging process, 
and spectral differences between the CMB and the sources used to measure the beam.  
The final uncertainties in the beam model parameters $\sigma_b$ and $a$ 
are constructed as the quadrature sum of the estimated uncertainties 
due to each of these individual sources of error.  
To a good approximation, the uncertainties on $\sigma_b$ and $a$ 
can be taken to be Gaussian and uncorrelated.

\begin{figure}[t]\centering
\includegraphics[width=0.5\textwidth]{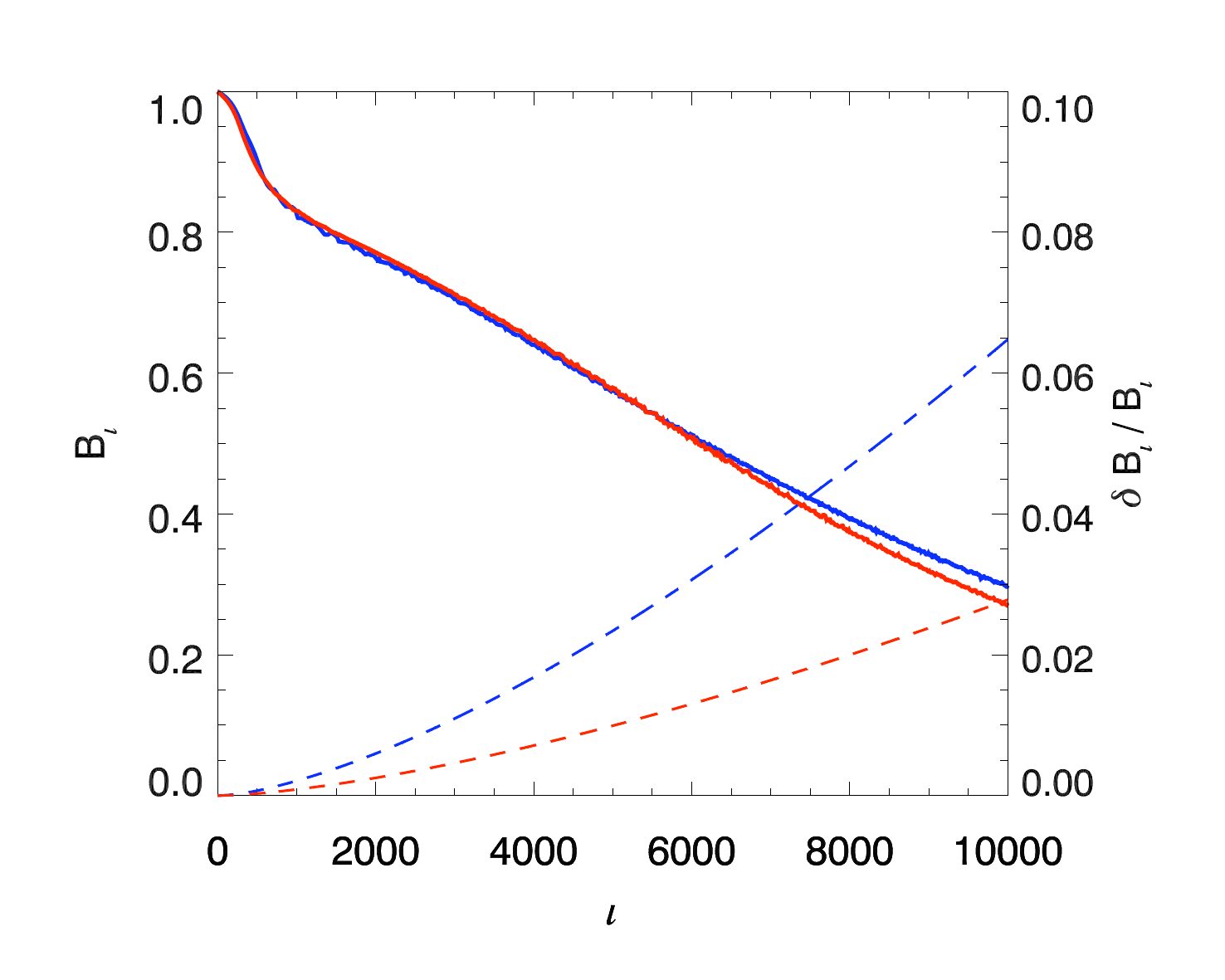}
  \caption[]{Average beam functions and uncertainties for SPT. 
  {\it Left axis:} The SPT beam function for $150\,$GHz ({\bf red}) and $220\,$GHz ({\bf blue}). 
  {\it Right axis:} The $1\,\sigma$ uncertainties on the beam function for each frequency.  
  The beam uncertainties shown here include only uncertainties on the main lobe beam width, $\sigma_b$, since the uncertainty of the sidelobe amplitude has been subsumed into the calibration
 uncertainty. }
\label{fig:beam}
\end{figure}

In practice, a change in the value of $a$ is equivalent to a change in
the overall calibration for $\ell > 700$.
After the beam uncertainties are
estimated, the uncertainty in $a$ is folded into the estimated
uncertainty on the absolute calibration and the
parameter $a$ is fixed to the best-fit value of 0.85. The quoted beam uncertainties in the {\textsc CosmoMC}\footnote{http://cosmologist.info/cosmomc} \citep{lewis02b} data files in \S\ref{sec:ps} include only the uncertainty on $\sigma_b$.
Figure \ref{fig:beam} shows the measured beam functions for 150 and
$220\,$GHz, along with the $1\,\sigma$ uncertainties in the main lobe beam
width $\sigma_b$. 
 
\subsection{Calibration}
\label{sec:calibration}

The calibration of the SPT data is tied to the superb WMAP5 absolute calibration through a direct comparison of $150\,$GHz SPT maps with WMAP5 V and W-band (61 and $94\,$GHz) maps \citep{hinshaw09} of the same sky regions.  
The WMAP5 maps are resampled according to the SPT pointing information, and the resulting TOD are passed through the SPT analysis pipeline to capture the effects of TOD filtering. 
The ratio of the cross-spectra of the filtered WMAP and SPT maps after correcting for the instrumental beams,
\begin{equation}
c = \frac{\left<a_{\ell m,{WMAP_i}}^{*}\,\,a_{\ell m,{WMAP_j}}\right>}{\left<a_{\ell m,SPT}^{*} \,\,a_{\ell m,{WMAP_j}} \left(\frac{B_\ell^{WMAP_i} }{B_\ell^{SPT}}\right)\right>},
\end{equation}
is used to estimate the relative calibration factor between the two experiments.
A similar procedure was used to calibrate the Boomerang, ACBAR, and QUaD experiments \citep{jones06,reichardt09a,brown09}.
Dedicated SPT calibration scans of four large fields totaling 1250~deg$^2$ of sky were obtained during 2008. 
The results for these four fields are combined to achieve an absolute temperature calibration uncertainty of 3.6\% at $150\,$GHz.

The $150\,$GHz calibration is transferred to $220$\,GHz through the overlapping coverage of SPT's high S/N maps. 
We calculate the relative calibration by examining the ratio of the cross-spectra between the 150 and $220\,$GHz maps to the auto-spectra of the $150\,$GHz map after correcting for the beam and filtering differences.  
We estimate the relative calibration uncertainty to be 6.2\% and the final absolute calibration uncertainty of the $220\,$GHz temperature map to be 7.2\%.

\section{Analysis}\label{sec:analysis}

In this section we describe the pipeline used to process the TOD to maps, and from maps to bandpowers. 
Given the small sky area analyzed here, we use the flat-sky approximation. 
We generate maps in a flat-sky projection and analyze these maps using Fourier transforms.  
Thus, the discussion of filtering and data-processing techniques refer to particular modes by their corresponding angular wavenumber {\bf k} in radians. Note that in these units $| {\bf k}| = \ell$.

\subsection{Map-making}
\label{sec:mapping}

We generate one map per frequency band for each two-hour observation of the field. 
The map-making pipeline used for this analysis is very similar to
that used by S09, with some small modifications.   An overview of the pipeline 
is provided below, emphasizing the aspects of data processing and
map-making that are most important for understanding the power spectrum
analysis.   

\subsubsection{Data Selection}
\label{sec:datasel}

Variations in observing conditions and daily receiver setup can affect the performance of individual
bolometer channels.  The first step in the data processing
is to select the set of bolometers with good performance for
each individual observation.  The criteria for selecting these
detectors are primarily based on responsivity (determined through the series of
calibrations performed prior to each field observation) and noise
performance. Detectors are rejected if they have a
low signal-to-noise response to a chopped thermal calibration source
or atmospheric emission (modulated through a short elevation scan).  They
are also rejected if their noise power spectrum is heavily contaminated by
readout-related line-like features.  More detail on the
bolometer cuts can be found in S09.  In addition, bolometers that have a responsivity or inverse-noise-based weight more than a factor of three
above or below the median for their observing band are
rejected.  
On average, 286 bolometers at $150\,$GHz (out of 394 possible) passed
all cuts for a given observation.  At $220\,$GHz, an average of 161 (out of 254 possible bolometers) passed all cuts.

In a given observation, segments of the TOD corresponding to individual scans may be rejected due to readout or cosmic-ray induced features, anomalous noise, or problems with pointing
reconstruction.  The scan-by-scan data cuts are the
same as those described in S09 and remove
approximately 10\% of the TOD.

The SPT detectors exhibit some sensitivity to the receiver's pulse-tube
cooler.  Bolometer noise power spectra are occasionally
contaminated by narrow spectral lines corresponding to the pulse-tube
frequency and its harmonics.  As in S09, we address this by applying a
notch filter to the affected data.  This filter removes less than 0.4\%
of the total signal bandwidth.  

In addition to the cuts above, we remove a small number of observations that have either incomplete 
coverage of the field or high noise levels.
Observations taken under poor atmospheric conditions are rejected in this way. 
The application of these cuts remove 36 out of the 336 total observations of the field.

\subsubsection{Time-Ordered Data (TOD) Filtering}
\label{sec:filtering}

Let $d_{\alpha j}$ be a measurement of the CMB brightness temperature
by detector $j$ at time $\alpha$.  Contributions to $d_{\alpha j}$ are the
celestial sky temperature, $s_{\alpha j}$, the atmospheric temperature,
$a_{\alpha j}$, and instrumental noise, $n_{\alpha j}$.  
The instrumental noise is largely uncorrelated between detectors.
However, the atmospheric signal is highly
correlated across the focal plane due to the overlap of individual detector beams as they pass through
turbulent layers in the atmosphere.  

The signals in the data have been low-pass filtered by the bolometers' optical time constants.
These single-pole bolometer transfer functions are measured
by the method described in S09.  We deconvolve the transfer functions
for each bolometer on a by-scan basis, as the
first step in TOD filtering.
At the same time, we apply a
low-pass filter with a cutoff at 12.5 Hz to remove noise above the
Nyquist frequency associated with the final  map pixelation. These
operations can be 
represented as linear operation ${\bf \Pi^{fft}}$ on the TOD,  
\begin{equation}
d'_{\alpha j} = \Pi^{\rm fft}_{\alpha \beta} d_{\beta j},
\end{equation}
where a sum is implied over repeated indices.

To remove 1/f noise and atmospheric noise in the scan direction, we project out a
19th order Legendre polynomial from the TOD for each detector on a
scan-by-scan basis.  This operation ${\bf \Pi^{t}}$ can be described by 
\begin{equation}
d''_{\alpha j} = \Pi^{\rm t}_{\alpha \beta} d'_{\beta j}.
\end{equation}
The effect on the signal is similar to having applied a
one-dimensional high-pass filter in the scan direction at $\ell \gtrsim 300$.   

Significant atmospheric signal remains in the data after this temporal
polynomial removal.  Because the atmospheric power is primarily  
common across the entire focal plane, we can exploit the spatial correlations to remove
atmosphere without removing fine-scale astronomical signal.  The method used in
this analysis is to fit for and subtract a plane, $a_0 + a_x {\bf x} +
a_y {\bf y}$, across all detectors in the detector array at each time
sample, where ${\bf x}$ and ${\bf y}$ are set by the angular
separation of each pixel from the boresight of the telescope.  
All detectors are equally weighted in this fit. This
operation can be represented as a linear operation on all detectors at
each time sample $\alpha$ to produce an atmosphere-cleaned dataset, 
${\bf d'''}$, 
\begin{equation}
d'''_{\alpha j} = \Pi^{\rm s}_{j k} d''_{\alpha k}.
\end{equation}

We mask the brightest point sources in the map before applying the
polynomial subtraction and the spatial-mode filtering.  The 92 sources
that were detected above 
$5\,\sigma$ in a preliminary $150\,$GHz map have been masked.  A
complete discussion of the point sources detected in this field can
be found in \citet[hereafter V09]{vieira09}.

\subsubsection{Map-making}

The data from each bolometer is inverse noise weighted according to the calibrated, pre-filtering detector PSD in the range 1-3~Hz, corresponding to $1400 < \ell <  4300$.
This multipole range covers
the angular scales where we expect the most significant detection of the SZ effect.
We represent the mapping between time-ordered bolometer samples and celestial positions with the pointing matrix  ${\bf L}$,  
which we apply to the cleaned TOD to produce a map ${\bf m}$, 
\begin{equation}
\label{eqn:mapgen}
{\bf m} = {\bf L  \Lambda_w \Pi^{\rm s}  \Pi^{\rm t} \Pi^{\rm fft} d}.
\end{equation}
${\bf  \Lambda_w}$ is a diagonal matrix encapsulating the detector weights. Information on the pointing reconstruction can be found in S09 and \cite{carlstrom09}.

\subsection{Maps to Bandpowers}
\label{sec:bandest}

We use a pseudo-$C_\ell$ method to estimate the bandpowers.  In pseudo-$C_\ell$ methods, bandpowers are estimated directly from the Fourier transform of the map after correcting for effects such as TOD filtering, beams, and finite sky coverage. 
We process the data using a cross spectrum based analysis \citep{polenta05, tristram05} in order to eliminate noise bias.  
Beam and filtering effects are corrected for according to the formalism in the MASTER algorithm \citep{hivon02}. 
We report the bandpowers in terms of $\mathcal{D}_\ell$, where
\begin{equation}
\mathcal{D}_\ell=\frac{\ell\left(\ell+1\right)}{2\pi} {\rm C}_\ell\;.
\end{equation}

As the first step in our analysis, we calculate the Fourier transform, $\tilde{m}_{\textbf{k}}$, of the single-observation maps for each frequency.  
Each map is apodized using the same window $\textbf{W}$, and thus $\tilde{m}^{(\nu,A)}_\textbf{k}\equiv\textrm{FT}\left(\textbf{W}\textbf{m}^{(\nu,A)}\right)$, where the first superscript, $\nu$, indicates the observing frequency, the second superscript, $A$, indicates the observation number, and the subscript denotes the angular frequency. 
Cross-spectra are computed for every map pair, and averaged within the appropriate $\ell$-bin $b$:
\begin{equation}
\label{eqn:ddef}
 D^{\nu_i\times\nu_j, AB}_b\equiv\frac{1}{N_b}\sum_{\textbf{k} \in b} \frac{{\rm k}({\rm k}+1)}{2\pi}\tilde{m}^{(\nu_i,A)}_\textbf{k} \tilde{m}^{(\nu_j,B)*}_\textbf{k}.
\end{equation}
We use the abbreviation, $D^{\nu_i}\equiv D^{\nu_i\times\nu_i}$, when referring to single-frequency auto-spectra. Recall that $| {\bf k} | = \ell$ in the flat-sky approximation.

The FFT of a map, $\widetilde{m}_\textbf{k}^{(\nu,A)}$, is linearly related to the FFT of the astronomical sky, $a_\textbf{k}^{\nu}$. It also includes a noise contribution, $n_\textbf{k}^{(\nu,Ai)}$, which is uncorrelated between the 300 observations:
\begin{equation}
\label{eqn:mvsa}
\widetilde{m}^{(\nu,A)}_\textbf{k}=\widetilde{W}_{\textbf{k}-\textbf{k}^\prime}G_{\textbf{k}^\prime}\left(
B_{\textbf{k}^\prime}a^{\nu}_{\textbf{k}^\prime}+n^{(\nu,A)}_{\textbf{k}^\prime}\right).
\end{equation}
Here $G_\textbf{k}$ is the 2-dimensional amplitude transfer function, which accounts for the TOD filtering as well as the map-based filtering described in \S\ref{sec:modeweighting}.  $\widetilde{\textbf{W}}_{\textbf{k}}$ is the Fourier transform of the apodization mask.

Following the treatment by \cite{hivon02}, we take the raw spectrum, $D^{AB}_b$, to be linearly related to the true spectrum by a transformation, $K_{bb^\prime}$.  
The $K$ transformation combines the power spectrum transfer function, $F_\ell$, which includes the effects of TOD-based and map-based filtering (\S\ref{sec:transfunccalc}), the beams, $B_\ell$ (\S\ref{sec:beams}), and the mode-coupling matrix, $M_{kk^\prime}[\textbf{W}]$ (\S\ref{sec:apodization}). The mode-coupling matrix accounts for the convolution of the spectrum due to the apodization window, $\textbf{W}$.
\begin{equation}
\label{eqn:kdef}
K^{\nu_i\times\nu_j}_{bb^\prime}=P_{bk}\left(M_{kk^\prime}[\textbf{W}]\,F^{\nu_i\times\nu_j}_{k^\prime}B^{\nu_i}_{k^\prime}B^{\nu_j}_{k^\prime}\right)Q_{k^\prime b^\prime}.
\end{equation}
Here, $P_{bk}$ is the re-binning operator \citep{hivon02}:
\begin{align}
P_{bk}&=\left \{\begin{array}{cl} \frac{k(k+1)}{2\pi \Delta k_{(b)}}&\;  k_{(b-1)} < k < k_{(b)}\\ 
0 & \textrm {otherwise} \end{array}\right.
\end{align}
while the inverse of the re-binning operator is $Q_{k b}$:
\begin{equation}
Q_{k b}=\left\{\begin{array}{cl}\frac{2\pi}{k(k+1)}&k_{(b-1)}< k <k_{(b)} \\ 0 &\textrm{otherwise}\end{array}\right. .
\end{equation}
In the rest of the paper, we will often refer to band averaged quantities, such as $C_b=P_{bk}C_k$.
Cross-spectra between observations that have been corrected for apodization and processing are denoted as, 
\begin{equation}
\widehat{D}^{\nu_i\times\nu_j,AB}_b\equiv \left(K^{-1}\right)_{bb^\prime}D^{\nu_i\times\nu_j,AB}_{b^\prime}.
\end{equation}
The final bandpowers are then computed as the average of all cross-spectra,
\begin{equation} 
\label{eqn:qdef}
q_b^{\nu_i\times\nu_j}=\frac{1}{n_\textrm{obs}\left(n_\textrm{obs}-1\right)}\sum_A \sum_{B \ne A} \widehat{D}^{\nu_i\times\nu_j,AB}_b .
\end{equation}

We make the simplifying assumption that the noise properties of each observation are statistically equivalent, hence the uniform weighting chosen here.  
The data selection criteria in \S\ref{sec:datasel} ensure that these observations target the same region of the sky,  have roughly the same number of active bolometers, and have similar noise properties.  
Therefore, this uniform weighting should be unbiased and only slightly sub-optimal.

The cross-spectrum bandpowers, $D^{\nu_i\times\nu_j,AB}_b$, generated from the 300 observations are used in conjunction with signal-only Monte Carlo bandpowers to generate empirical covariance matrices, as described in appendix \ref{app:covmatrix}.   
The variance of the Monte Carlo bandpowers is used to estimate the sample variance contribution to the covariance matrix.  
Meanwhile, we use the variance in the spectra of the real data to estimate the uncertainty due to noise in the maps.

\subsubsection{Apodization Mask and Calculation of the Mode-mixing Kernel} 
\label{sec:apodization}
Since we have only mapped a fraction of the full sky, the angular power spectra of the maps are convolutions of the true  $C_{\ell}$s with an $\ell$-space, mode-mixing kernel that depends on the map size, apodization, and point source masking.
We calculate this mode-mixing kernel, $M_{kk^\prime}[\textbf{W}]$, following the derivation in \cite{hivon02} for the flat sky case:  

\begin{equation}
M_{kk^\prime}[\textbf{W}]\equiv\frac{1}{(2\pi)^2}\int d\theta_k\, d\theta_{k^\prime} \left|\widetilde{W}_{\textbf{k}-\textbf{k}^\prime}\right|^2.
\end{equation}  
If the mask is smooth on fine angular scales, then the mode coupling kernel can be approximated by a delta function at high-$k$:  
\begin{equation}
\label{eqn:mnorm}
M_{kk^\prime}[\textbf{W}]\approx w_2 \delta_{kk^\prime}.
\end{equation}
Here we use the notation 
$w_n\equiv\left<\textbf{W}^n\right>$ 
to represent the $n^\textrm{th}$ moment of the apodization window.   
In this limit, the coupling kernel serves the purpose of re-normalizing the power spectrum to account for modes lost due to apodization.  
As we discuss in appendix \ref{app:covmatrix}, the coupling kernel also plays an important role in determining the shape and normalization of the covariance matrix.

We avoid areas of the map with sparse or uneven coverage in any single observation. Thus, the apodization window is conservative in its avoidance of the map edges.
We also mask 144 point sources detected in the $150\,$GHz data above a significance of $5\,\sigma$, which corresponds to a flux of $6.4\,$mJy.   This source list is from a more refined analysis than the preliminary list used in \S\ref{sec:filtering}; the differences are in sources near the $5\,\sigma$ detection threshold. Each point source is masked by a 2'-radius disk, with a Gaussian taper outside this radius.    
Many different tapered mask shapes were tested for both efficacy in removing point source power and noise performance.  Given the relatively small area masked by point sources, varying the shape of the point source mask has little effect on the final spectrum.
The effective sky area of the mask is$78\,$deg$^2$.
Simulations have been performed to test whether the application of this mask will bias the inferred 
power and we see no bias with a $5\,\sigma$ cut. 
As could be expected, we do observe a mild noise bias when using a more
aggressive $3\,\sigma$ point source mask.  The same $5\,\sigma$ mask is used for all maps at both frequencies and for all observations.

\subsubsection{Fourier Mode Weighting} 
\label{sec:modeweighting}
The maps produced  by the steps described in \S\ref{sec:filtering} have anisotropic signal and noise. 
In particular, the map noise PSD rises steeply at spatial frequencies corresponding to low spatial
frequencies in the scan direction (low $k_x$). 
The covariance of the power estimated in a given $\ell$-bin depends on the second power of the map noise PSD for all the Fourier modes in that $\ell$-bin.  In the presence of either non-uniform noise or signal, applying an optimal mode-weighting when calculating the mean bandpower may significantly reduce the final noise covariance matrix of the power spectrum bandpowers. 
In the case of SPT, we find that for the purposes of measuring the $\ell \gtrsim 2000$ power spectrum,
a simple, uniform selection of modes at $k_x > 1200$ is close to optimal, 
and we apply this mode weighting in calculating the SPT bandpowers.

\subsubsection{Transfer Function Estimation} 
\label{sec:transfunccalc}

 In order to empirically determine the effect of both the TOD-based filtering and the Fourier mode-weighting, we calculate a transfer function, $F_k$, as defined in \cite{hivon02}.   Note that this  power spectrum transfer function is distinct from the amplitude transfer function $G_{\bf k}$.  
 Specifically, the transfer function accounts for all map- and TOD-processing not taken into account by the mode-coupling kernel or the beam functions.

\label{sec:simulations}

We created 300 Monte Carlo sky realizations at 150 and 220$\,$GHz in order to calculate the transfer function of the filtering. 
The simulations also serve as an input for the covariance matrix estimation.  
These simulations contain two components: a CMB component and a point source component.   
The CMB component is computed for the best-fit WMAP5 lensed $\Lambda$CDM model.

The point source component includes two different populations of dusty galaxies, a low-z population and a high-z population. 
For each population we generate sources from a Poisson distribution.  
Sources are generated in bins of flux, $S$,  ranging from 0.01 to $6.75\,$mJy. 
This upper limit in flux is close to the $5\,\sigma$ detection threshold in the $150\,$GHz maps. 
In each flux bin, the $150\,$GHz source density, $dN/dS$, of each population is taken  from the model of \citet{negrello07}.  
We relate the flux of each source at $220\,$GHz to its flux at $150\,$GHz with 
a power law in intensity, $S_\nu \propto \nu^\alpha$.
The power law spectral index, $\alpha$, for each source is drawn from a normal distribution.
We use spectral indices of $\alpha=3\pm0.5$ for the high-z protospheroidal galaxies, and $\alpha=2\pm0.3$ for the low-z IRAS-like galaxies. 
As with any Poisson distribution of point sources, the power spectrum of these maps is white ($C_\ell^{\rm ps}$ = constant) and related to the flux cutoff, $S_0$, by:
\begin{equation}
C_\ell^{\rm ps} = \left(\frac{dB_\nu}{dT}\right)^{-2}_{\rm T_{\rm CMB}}\int_0^{S_0} S^2 \frac{dN}{dS} dS.
\end{equation}
The power spectra of these simulated point source maps are well represented by a constant $C_\ell^{\rm ps} = 1.1 \times 10^{-5}\,\mu{\rm K^2}$ at $150\,$GHz and $C_\ell^{\rm ps} = 6.8\times 10^{-5}\,\mu{\rm K^2}$ at $220\,$GHz.

These simulated maps are smoothed by the appropriate beam.   
From each map realization, we construct simulated TOD using the pointing information.   
Each realization of the TOD is processed using the low-pass filter, polynomial removal, and the spatial-mode subtraction described in \S\ref{sec:filtering}.    
No time-constant deconvolution is applied, since these realizations of the TOD  have not been convolved by the bolometer time constants.  The filtered, simulated TOD are then converted into maps according to equation \ref{eqn:mapgen}.

We apply the same 
apodization mask and Fourier mode weighting to these map realizations as is used on the actual data. 
We then compute the Monte Carlo pseudo-power spectrum, $\left(D_k\right)_{\rm MC}$, for each map.  
The simulated transfer function is calculated iteratively, 
by comparing the Monte Carlo average, $\left<D_k\right>_\textrm{MC}$, to the input 
theory spectrum, $C_k^{\textrm{theory}}$. 

For the single-frequency bandpowers, we start with an initial guess of the transfer function:  
\begin{equation}
\label{eqn:transguess}
F^{\nu,(0)}_{k}=\frac{\left<D^{\nu}_{k}\right>_\textrm{MC}}{w_2 {B^{\nu}_k}^2 C^{\nu,\textrm{theory}}_k}.
\end{equation}
The superscript, $(0)$, indicates that this is the first iteration in the transfer function estimates.
 For this initial guess, we approximate the coupling kernel as largely diagonal as in equation \ref{eqn:mnorm}. 
Thus, the factor $w_2$ is the normalization factor required by the apodization window.
We then iterate on this estimate using the full mode-coupling kernel:
\begin{equation}\label{eqn:transiter}
F^{\nu, (i+1)}_{k}=F^{\nu,(i)}_{k}+\frac{\left<D^{\nu}_{k}\right>_{\rm MC} -  M_{kk^\prime} F_{k^\prime}^{\nu,(i)} {B^\nu_{k^\prime}}^2 C^{\nu,\textrm{theory}}_{k^\prime}}{{B^\nu_{k}}^2 C^{\nu,\textrm{theory}}_{k} w_2}.
\end{equation}
We iterate on this estimate five times, although the transfer function has largely converged after the first iteration.   

This method may  misestimate the transfer function if the simulated spectrum is significantly different from the true power spectrum.
The primary CMB anisotropy has been adequately constrained by previous experiments, however, the foreground power spectrum is less well known.  
We repeated the transfer function estimation using an input power spectrum with twice the nominal point source power.     
The resulting transfer function was unchanged at the 1\% level, giving confidence that the transfer function estimate is robust.  

The transfer function for a multifrequency cross-spectrum is taken to be the geometric mean of the two individual transfer functions: 
\begin{equation}
F_k^{\nu_i\times\nu_j}=\sqrt{F_k^{\nu_i}F_k^{\nu_j}}.
\end{equation}
This treatment is only strictly correct for isotropic filtering.  
  As a cross check, we have also computed the cross-spectrum transfer function directly. 
For the angular multipoles reported in this work, the geometric-mean transfer function estimate is in excellent agreement with the estimate obtained using $\left<D^{\nu_1\times\nu_2}_k\right>_{\rm MC}$ in equations \ref{eqn:transguess} and \ref{eqn:transiter}.

\subsubsection{Frequency-differenced Spectra}

\label{subsec:freqsub}

We are interested in the power spectra of linear combinations of the $150$ and 220$\,$GHz maps designed to remove astronomical foregrounds.
One method for obtaining such power spectra would be to directly subtract the maps
after correcting the maps for differences in beams or processing.    
In such a map subtraction, the scaling of the $220\,$GHz map, $x$, can be adjusted to optimally remove foregrounds.  
The differenced maps can then be processed using our standard pipeline to obtain spectra with a reduced foreground contribution.  

Equivalently, the differenced spectrum can be generated from the original 
bandpowers, $q_b^i$, using a linear spectrum transformation, $\xi$:   
\begin{equation}
q^{\textrm{150}-x\times\textrm{220}}_b=\sum_{i} \xi^{i}(x) q_b^i.
\end{equation}
\noindent Here, the index, $i$, denotes the $150\,$GHz auto-spectrum, $220\,$GHz auto-spectrum, or $150\times 220\,$GHz cross spectrum.  
This transformation is computationally fast and takes advantage of the fact that the cross-frequency bandpowers include information on the relative phases of each Fourier component in the map.
In this way, the difference spectrum can be represented in terms of the three measured spectra:
\begin{align}
\notag q_b^{\nu_i-x\times\nu_j} &=\frac{1}{\left(1-x\right)^2}\sum_{\textbf{k}\in b} | a_\textbf{k}^{\nu_i}-x a_\textbf{k}^{\nu_j}|^2\\
\notag&=\frac{1}{\left(1-x\right)^2}\sum_{\textbf{k} \in b}|a_\textbf{k}^{\nu_i}|^2-2x \textrm{Re}\left(a_\textbf{k}^{\nu_i} a_\textbf{k}^{\nu_j*}\right)+x^2 |a_\textbf{k}^{\nu_j}|^2\\
\label{eqn:cldiff}&=\frac{1}{\left(1-x\right)^2}\left(q_b^{\nu_i}-2 x q_b^{\nu_i \times \nu_j}+x^2 q_b^{\nu_j}\right).
\end{align}
The overall normalization is chosen such that the CMB power is unchanged in the subtracted spectrum.
For clarity, we have momentarily avoided here the complications of beams and filtering and have expressed the bandpowers in terms of the Fourier transform of the celestial sky, $a_\textbf{k}$.
For a given proportionality constant, $x$, the values of $\xi(x)$ are:
\begin{eqnarray}
\xi^{\textrm{150}}(x)&=&\frac{1}{(1-x)^2}, \nonumber\\
\xi^{\textrm{150}\times\textrm{220}}(x)&=&\frac{-2x}{(1-x)^2},\nonumber\\
\xi^{\textrm{220}}(x)&=&\frac{x^2}{(1-x)^2}.
\end{eqnarray}
\noindent We also compute the covariance matrix for the subtracted spectrum from the original bandpower covariance matrices:
$$\textbf{C}^{150-x\times220}_{bb^\prime}=\sum_{i,j} \xi^{i}(x)\textbf{C}^{(i,j)}_{b b^\prime} \xi^{j}(x) .$$

\section{Jack-knives}
\label{sec:jackknives}

We apply a set of jack-knife tests to the SPT data to search for
possible systematic errors. In a jack-knife test, the data set is
divided into two halves,
based on features of the data associated with potential sources of systematic error.
The two halves are differenced to remove any astronomical
signal, and the resulting power spectrum is compared to
zero. Significant deviations from zero would 
indicate either a systematic problem or a noise misestimate.  
We implement the jack-knives in the cross-spectrum framework by differencing single pairs of observations and applying the cross-spectrum estimator outlined in \S\ref{sec:bandest} to the set of differenced pairs.   
In total, we perform 13 different jack-knife tests.    

\begin{figure*}[ht]\centering
\includegraphics[width=0.8\textwidth]{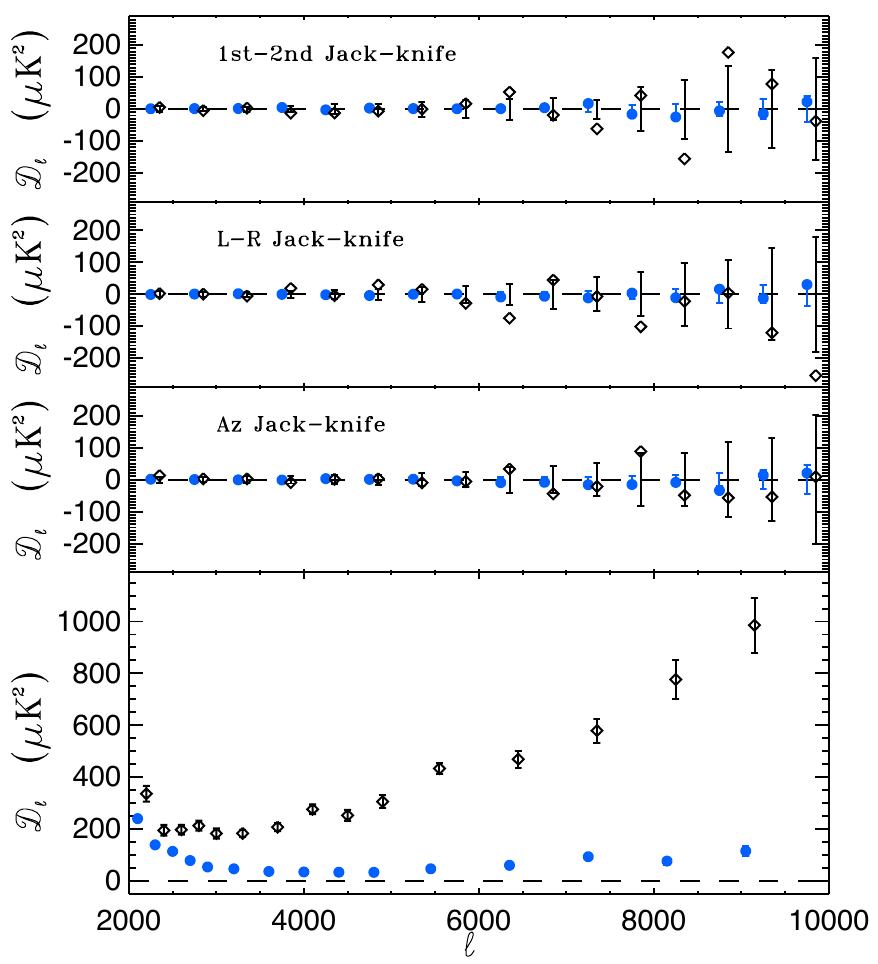}
  \caption[]{ Jack-knives for the SPT data set at $150\,$GHz ({\bf blue circles}) and $220\,$GHz ({\bf black diamonds}). 
  For clarity, the $220\,$GHz jack-knives have been shifted to the right by $\Delta\ell = 100$. 
  {\it Top panel:} Bandpowers of the ``first half - second half" jack-knife compared to the expected error bars about zero signal. 
  Disagreement with zero would indicate either a noise misestimate or a time-dependent systematic signal.   
  {\it Second panel:} Power spectrum of the left-going minus right-going difference map. 
  This test yields strong constraints on the accuracy of the detector transfer function deconvolution and on possible directional systematics. 
  {\it Third panel:} Bandpowers for the difference map when the data is split based on azimuth. 
  Signals fixed in azimuth such as side-lobe pickup from the nearby support building would produce non-zero power.  
  We see no evidence for ground-based pickup. The cumulative probability to exceed the $\chi^2$ observed in these three tests is 76\% at 150~GHz and 22\% at 220~GHz. {\it Bottom panel:} The un-differenced SPT power spectra at each frequency for comparison.}
\label{fig:jack}
\end{figure*}

Six jack-knives are based on the observing parameters, such as time, scan direction and azimuthal range.
The data can be split based on when it was taken to search for
systematic changes in the calibration, beams, detector time constants,
or any other time variable effect. 
The ``first half - second
half" jack-knife probes variations on month time scales, while an
``even\,-\,odd" jack-knife differencing every other observation looks for
variations on hourly time scales. 
Results for the ``first half - second
half" jack-knife are shown in the top panel of Figure \ref{fig:jack}. The data can also
be split based on the direction of the scan in a ``left\,-\,right"
jack-knife (panel 2 of Figure \ref{fig:jack}). We would expect to see
residual power here if the detector transfer
function had been improperly de-convolved, if the telescope
acceleration at turn-arounds induces a signal through sky modulation or microphonics, or if
the wind direction is important. We observed this field at two scan
speeds different by 10\%.  We check for systematic differences
related to the scan speed, such as a mirror wobble, in a ``scan speed"
jack-knife.  Side-lobe pickup could potentially introduce spurious
signals into the SPT maps from the moon or features on the ground. We test for
moon pickup by splitting the data based on whether the moon is above or below
the horizon.  We test for ground pickup by splitting on the mean
azimuth of the observation. To maximize 
the sensitivity to ground pickup, the azimuthal ranges are selected to be
centered on and directly opposite the closest building to
the telescope, which is the most likely source of ground
signal. The azimuthal jack-knife is shown in the third panel of 
Figure \ref{fig:jack}.  

We also perform jack-knives based on four
noise and observation-quality statistics of the $150\,$GHz data.  
The first is based on the overall RMS in
the maps, which is affected by atmospheric conditions and detector
noise. The second is based on the average raw detector PSD in the
range 9-11 Hz,  which is a measure of the detector ``white''
noise level.  The third is the RMS near $\ell=3000$
where the S/N on the SZ power spectrum is highest.   The fourth is based on 
the number of bolometers active in each observation.

There are also a number of line-like spectral
features in the SPT TOD that could potentially affect the
power spectrum bandpowers, and we perform three jacknife tests for sensitivity to these features.  
Some of these line features appear at harmonics of the receiver 
pulse-tube  frequency, and are typically correlated across many bolometer
channels. These lines are filtered from the data as described in \S\ref{sec:filtering}.
In addition, some channels exhibit occasional line-like features at
other frequencies, which are not filtered in the data processing.
We search for residual effects in a jack-knife based on the average number of
line-like features for all $150\,$GHz bolometers, as well as a jack-knife based on the 
bandwidth affected by the lines.  Finally, we do an additional split using 
the average number of lines in an observation that appear to be related 
to the pulse-tube cooler.

We calculate the $\chi^2$ with respect to zero for
each jack-knife over the range $\ell \in [2000,10000]$ in bins with $\Delta\ell = 500$.
Some of the tests are highly correlated. For example, we changed scan
velocities approximately midway through the observations so splitting
the data based on scan velocity is nearly identical to splitting the
data between the first and second halves of the season. We calculate a
correlation coefficient between the different tests by adding
$1/N_{obs}$ for each common observation in a half, and subtracting
$1/N_{obs}$ for each distinct observation in a half. This algorithm
returns unity for two identical sets and zero for two random sets, as
we expect 50\% of the observations to be in common for two random
selections. The correlation coefficients between the 13 jack-knives
range from 0 to 0.83, with the maximum correlation being for the
previously mentioned scan velocity and first half - second half
jack-knives. 
% Comment to typesetter:
% \mathcal{C} used here to distinguish the jack-knife correlation matrix
% from the usual covariance matrix {\bf C}
We invert the jack-knife correlation matrix, ${\bf \mathcal{C}}$, and calculate
$\chi^2 = v_{i\alpha} ({\mathcal{C}}^{-1})_{ij} v_{j\alpha}$. 
Here $v_{i\alpha}$ is the ratio of the bandpower over the uncertainty for the ${i}^{\rm th}$ jack-knife and ${\alpha}^{\rm th}$ $\ell$-bin. 
The probability to exceed the measured $\chi^2$ for the
complete set of thirteen jack-knives is 77\% for the $150\,$GHz data, 32\% 
for the $220\,$GHz data and 57\% for the combined set with both
frequencies.  We thus find no evidence for systematic contaminants in the
SPT data set. 

\section{Foregrounds} 
\label{sec:foregrounds}

The main foregrounds at frequencies near $150\,$GHz are expected to be galactic dust emission, radio sources, and dusty star forming galaxies (DSFGs). 
Note, however, that the SPT field is selected to target one of the cleanest regions on the sky for galactic dust emission, and in the \citet{finkbeiner99} model, dust emission is primarily on large angular scales. 
The contribution for the selected field on arcminute scales is insignificant.  
The primary foregrounds of consideration for this analysis are radio sources and DSFGs.

Tens of bright radio sources are detected in the SPT maps at $>5\,\sigma$, and contribute substantial amounts of power at both 150 and $220\,$GHz.   Information on the fluxes and spectral indices of these and other sources significantly detected in the SPT maps can be found in V09.   Without masking, the measured point source power is $C_\ell^{\rm unmasked} = 2.1\times10^{-4} \,\mu{\rm K}^2$ at 150~GHz  and $C_\ell^{\rm unmasked} = 1.6\times10^{-4} \,\mu{\rm K}^2$ at 220~GHz. These estimates are dominated by the brightest few sources and thus subject to very large sample variance.
We mask all sources with $150\,$GHz fluxes above the $5\,\sigma$ detection threshold, $S=6.4\,$mJy.  
By masking these sources we reduce the radio source contribution to the SPT bandpowers by several orders of magnitude.  
According to the \citet{dezotti05} model source counts, after masking bright sources, we expect a residual radio source contribution of $C_\ell^{\rm radio} = 8.5\times10^{-7} \,\mu{\rm K}^2$ at $150\,$GHz.

The point source masking will remove the SZ contribution from only a few galaxy clusters, leading to negligible reduction of the SZ power. 
This is because the large majority of radio sources and DSFGs reside outside of SZ clusters.
Most of the masked sources are identified as radio sources in V09. 
Extrapolations from lower frequency observations imply that, at $150\,$GHz, less than 3\% 
of clusters contain radio source flux exceeding 20\% of the tSZ flux decrement \citep{lin09,sehgal09}. 
The masked sources are selected as increments at $150\,$GHz and therefore have fluxes much greater than 20\% of the tSZ of any associated galaxy cluster.
The number of clusters masked by the radio source masking should then be much 
less than 3\% and negligible. 
We also compare the tSZ power spectrum in the \citet{sehgal09} simulated sky maps with and without masking $> 6.4\,$mJy sources and find the difference to be $<<$ 1\%. 
A small number (six) of the masked sources are identified as DSFGs (V09).  
Galaxy clusters have a DSFG abundance only twenty times larger than the field \citep{bai07}, although they exceed the mass density of the field by a factor of 200 or more. 
Given the relative rarity of galaxy clusters, it follows that only a small fraction of DSFGs can live 
in galaxy clusters.
Therefore, the number of clusters masked along with the DSFGs should be much smaller than six and negligible.

Both the DSFG and radio source arguments above depend implicitly on 
the impact of potentially masking $\ltsim\,10$ clusters. 
As a worst-case study, 
\citet{shaw09} consider the impact of masking the most massive ten clusters in the field and show that even in this extreme case, the tSZ power spectrum at $\ell = 3000$ is reduced by only 11\%.  
Of course, the point source masking will not select the most massive clusters and is highly unlikely to remove as many as ten clusters. Hence, the true impact will be significantly less.

After we mask sources above the $5\,\sigma$ threshold, DSFGs are the dominant point source population in the SPT maps.  
These sources have been extensively studied at higher frequencies by SCUBA \citep{holland99}, MAMBO \citep{kreysa98}, Bolocam \citep{glenn98}, 
LABOCA \citep{siringo09}, 
AzTEC \citep{scott08}, SCUBA-2 \citep{holland06}, and BLAST \citep{pascale08}, 
and there have been some preliminary indications of their contribution in previous small-scale power spectra at $150\,$GHz \citep{reichardt09a,reichardt09b}. 
The flux of these galaxies has been observed to scale to higher frequencies as $S_\nu \propto \nu^{2.4-3.0}$ \citep{knox04,greve04,reichardt09b}, with the exact frequency dependence a function of the dust emissivity, the dust temperature, and the redshift distribution of the galaxy population. 
This range of
spectral indices corresponds to point source amplitude ratios, $\delta T^{\rm ps}_{ 220}/\delta T^{\rm ps}_{ 150}$, of 2.1-2.6 when
expressed in units of CMB temperature.
The measured spectral index of the DSFGs in the SPT maps is discussed extensively in H09. 

In order to obtain an unbiased estimate of the SZ power spectrum, it is essential that we take these sources into account in our fits and modeling.
After masking the bright point sources, we significantly detect a Poisson distributed 
power at $150\,$GHz of $C_\ell^{\rm ps} =7.1 \pm 0.5 \times 10^{-6} \mu{\rm K}^2$ (H09). 
This unclustered point source power climbs with increasing $\ell$ to be comparable to the SZ effect by $\ell = 2500-3000$, and is the dominant astronomical signal in the maps at arcminute scales.

The distribution of DSFGs on the sky is also expected to be clustered, resulting in a significant increase in power at $\ell \lesssim 3000$. BLAST recently detected this clustered term for DSFGs at 600 - 1200$\,$GHz (500 - 250$\,\mu$m) \citep{viero09}. 
Extrapolating the measured clustering to 150$\,$GHz, we expect the clustered contribution to be comparable to the tSZ effect. 
H09 have analyzed the SPT 150, 220, and $150\times220\,$GHz bandpowers presented in \S\ref{sec:results}, and have also found that the amplitude of the clustered component is indeed comparable to the tSZ amplitude. 
Discriminating between clustered DSFGs and the tSZ effect would be extremely difficult for a single-frequency instrument as the angular dependencies are very similar. However, the two frequencies used in this analysis allow the spectral separation of the these two astronomical signals.

\section{Results}
\label{sec:results}

In addition to presenting the SPT angular power spectra, we wish to determine the amplitude of the tSZ power spectrum and to use this amplitude to constrain the normalization of the matter power spectrum, $\sigma_8$. 
This measurement requires separating the tSZ signal from the other astrophysical signals in our data which 
include primary CMB anisotropy (including lensing effects), DSFGs (both Poisson and 
clustered components), and anisotropy due to the kSZ effect.    
The primary CMB anisotropy and Poisson point source component can be separated from 
a tSZ-like component on account of the distinct angular power spectra of these three signals.  
However the tSZ, kSZ, and clustered DSFG components are all expected to be roughly flat in $\mathcal{D}_\ell$, and we depend on their distinct frequency dependences to separate them.
We use a combination of the two frequency bands to remove the DSFG contribution to the power spectra.
We address the remaining degeneracy between the tSZ and kSZ effects by repeating the analysis for a range of assumed kSZ models. 

In \S\ref{sec:ps} we present bandpowers for 150, 220 and \nolinebreak{$150\times220\,$GHz} SPT angular power spectra.  
In \S\ref{sec:ptsrcsub} we combine the two frequency bands to generate a set of DSFG-subtracted bandpowers.   
In \S\ref{sec:params} we describe the Monte Carlo Markov chain (MCMC) analysis used to estimate the tSZ power spectrum amplitude, parametrized as the normalization $A_{\rm SZ}$ of a model template, from the DSFG-subtracted bandpowers.   
In \S\ref{sec:sigma8} we discuss the implications of the measured tSZ power spectrum amplitude for $\sigma_8$ and modeling of the tSZ effect.

\subsection{Power Spectra}\label{sec:ps}

\begin{figure*}[ht]\centering
\includegraphics[width=0.9\textwidth]{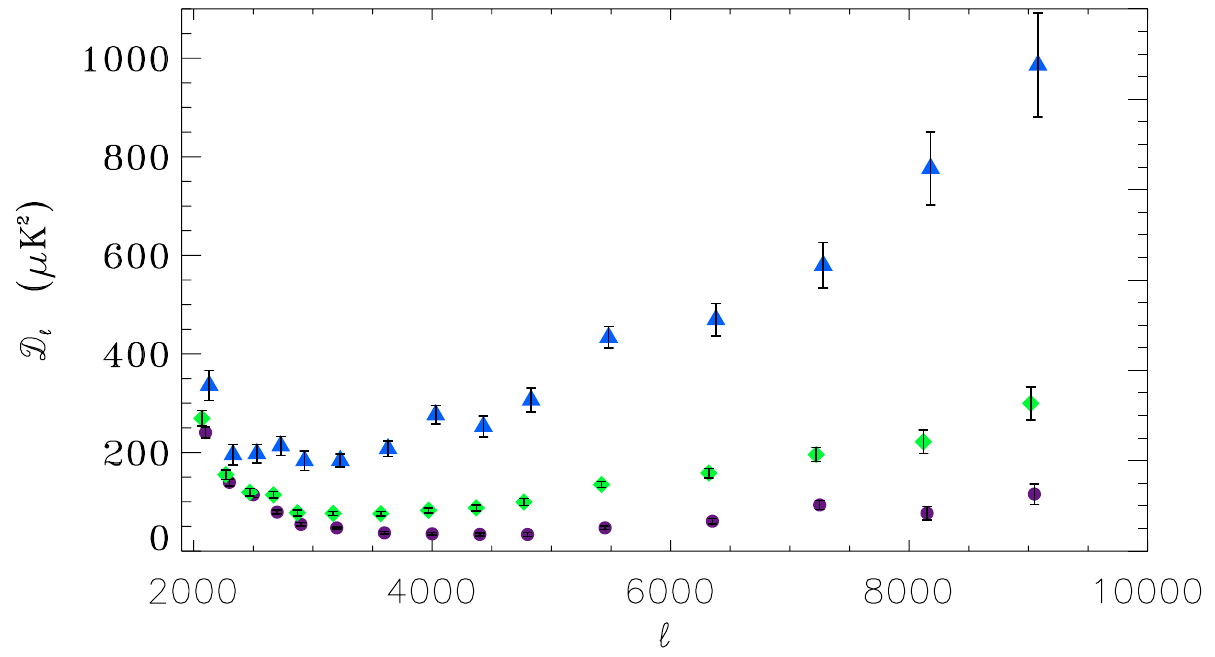}
  \caption[]{The SPT $150\,$GHz ({\bf purple circles}), $150\times220\,$GHz ({\bf green diamonds}) and $220\,$GHz ({\bf blue triangles}) bandpowers.
  The damping tail of the primary CMB anisotropy is apparent below $\ell = 3000$. 
  Above $\ell = 3000$, there is a clear excess with an angular dependence consistent with point sources.  
  These sources have low flux (as sources with $>6.4\,$mJy at $150\,$GHz have been masked) and a rising frequency spectrum, consistent with our expectations for Poisson distributed DSFGs.  
 The point source population and resulting contributions to anisotropy power are discussed in more detail in H09.}
\label{fig:clspt}
\end{figure*}

\begin{figure*}[ht]\centering
\includegraphics[width=0.9\textwidth]{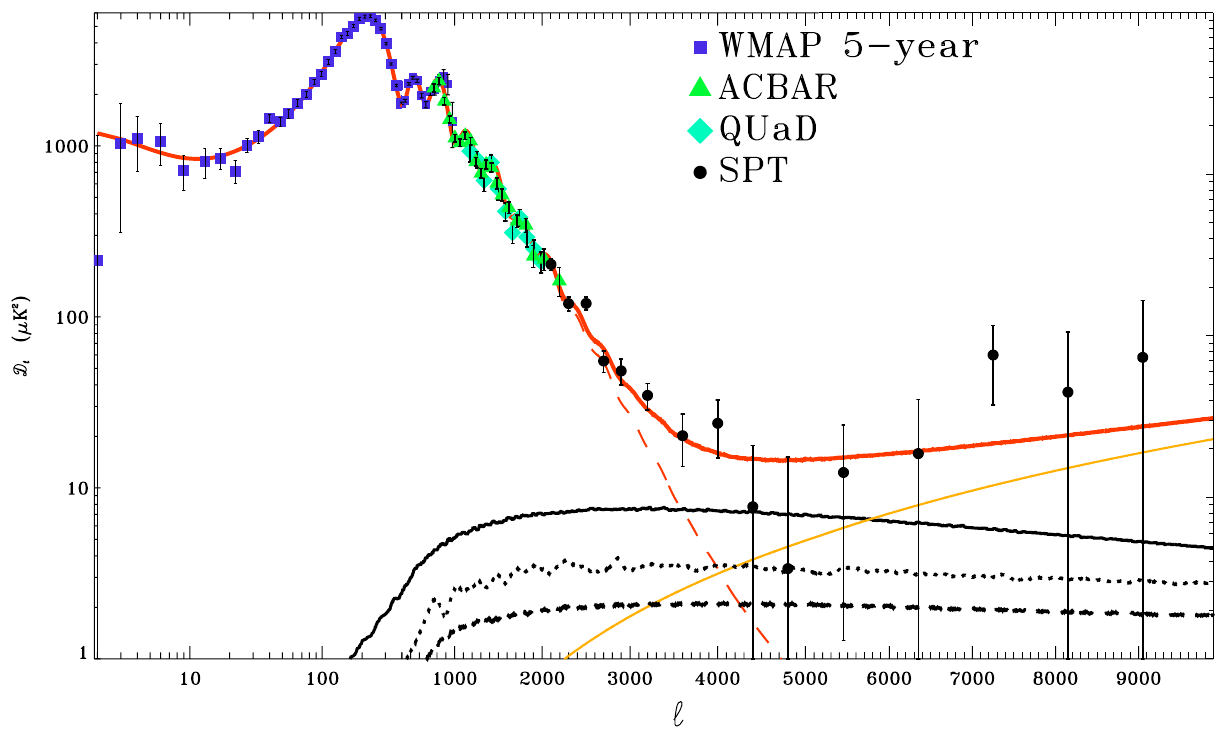}
  \caption[]{WMAP5 ({\bf blue squares}), ACBAR ({\bf green triangles}), QUaD ({\bf turquoise diamonds}) and the SPT ({\bf black circles}) DSFG-subtracted SPT bandpowers are plotted over the best-fit, lensed $\Lambda$CDM cosmological model ({\bf dashed red line}), best-fit tSZ power spectrum ({\bf solid black line}), homogeneous kSZ model ({\bf dashed black line}), and residual Poisson-distributed point source contribution ({\bf solid orange line}).
   The combined best-fit model is shown by the {\bf solid red line}. The plotted SPT bandpowers have been multiplied by the best-fit calibration factor of 0.92.
  Point sources above $6.4\,$mJy at $150\,$GHz have been masked. 
  The patchy kSZ template is also shown for reference ({\bf dotted black line}).   The DSFG-subtracted bandpowers are normalized to preserve the amplitude of the primary CMB anisotropies.
 }
\label{fig:clall}
\end{figure*}

Figure~\ref{fig:clspt} shows the bandpowers we compute by applying the analysis methods described in \S\ref{sec:analysis} to one 100 deg$^2$ field observed by SPT at 150 and $220\,$GHz. 
The bandpowers for the two frequencies and their cross-spectrum are tabulated in Table \ref{tab:bandpowerssinglefreq}. 
We combine these multi-frequency data as described in \S\ref{sec:analysis} to produce the `DSFG-subtracted' bandpowers listed in Table~\ref{tab:bandpowers}. 
This power spectrum is compared to the results from WMAP5, ACBAR, and QUaD in Figure~\ref{fig:clall}. 
The best-fit model to this combined data set including the primary CMB, kSZ, tSZ, and a Poisson point source contribution is shown for reference. The primary CMB anisotropy is estimated for a 
spatially-flat, $\Lambda$CDM model, which includes gravitational lensing. 
The bandpower uncertainties are derived from the combination of simulations and the measured intrinsic variations within the SPT data as described in \S\ref{sec:analysis}. 
The bandpowers can be compared to theory using the associated window functions \citep{knox99}. 
The bandpowers, uncertainties, and window functions may be downloaded from the SPT website.\footnote{http://pole.uchicago.edu/public/data/lueker09/}

\begin{table*}[ht!]
\begin{center}
\caption{\label{tab:bandpowerssinglefreq} Single-frequency bandpowers}
\small
\begin{tabular}{cc|cc|cc|cc}
\hline\hline
\rule[-2mm]{0mm}{6mm}
& &\multicolumn{2}{c}{$150\,$GHz} & \multicolumn{2}{c}{$150\times220\,$GHz} & \multicolumn{2}{c}{$220\,$GHz} \\
$\ell$ range&$l_{\rm eff}$ &$q$ ($\mu{\rm K}^2$)& $\sigma$ ($\mu{\rm K}^2$) &$q$ ($\mu{\rm K}^2$)& $\sigma$ ($\mu{\rm K}^2$)&$q$ ($\mu{\rm K}^2$)& $\sigma$ ($\mu{\rm K}^2$)\\
\hline
2001 - 2200 & 2058 & 240.5 &  11.4 & 269.5 &  15.3 &  336.2 &  30.4 \\ 
2201 - 2400 & 2276 & 139.4 &   7.0 & 155.0 &   9.8 &  195.4 &  20.5 \\ 
2401 - 2600 & 2474 & 114.3 &   5.2 & 119.2 &   8.1 &  197.6 &  18.8 \\ 
2601 - 2800 & 2677 &  79.0 &   4.3 & 114.0 &   6.7 &  213.3 &  19.2 \\ 
2801 - 3000 & 2893 &  54.1 &   3.7 &  77.7 &   6.1 &  183.1 &  19.5 \\ 
3001 - 3400 & 3184 &  47.0 &   2.4 &  76.5 &   4.2 &  183.7 &  13.4 \\ 
3401 - 3800 & 3581 &  36.9 &   2.4 &  75.8 &   4.3 &  207.7 &  15.6 \\ 
3801 - 4200 & 3992 &  35.0 &   2.8 &  82.7 &   5.0 &  276.5 &  18.8 \\ 
4201 - 4600 & 4401 &  33.9 &   3.2 &  87.5 &   5.9 &  252.8 &  21.0 \\ 
4601 - 5000 & 4789 &  33.6 &   4.2 &  99.4 &   7.3 &  306.1 &  24.5 \\ 
5001 - 5900 & 5448 &  47.1 &   3.5 & 135.1 &   6.3 &  433.6 &  21.7 \\ 
5901 - 6800 & 6359 &  60.5 &   5.5 & 158.2 &   9.4 &  469.6 &  33.0 \\ 
6801 - 7700 & 7255 &  93.5 &   8.8 & 195.8 &  14.5 &  579.7 &  46.4 \\ 
7701 - 8600 & 8161 &  76.8 &  13.5 & 221.8 &  23.6 &  776.5 &  73.9 \\ 
8601 - 9500 & 9059 & 115.6 &  20.7 & 299.7 &  33.7 &  986.1 & 105.9 \\ 
\hline
\end{tabular}
\tablecomments{ Band multipole range and weighted value $\ell_{\rm eff}$, bandpower $q_B$, 
and uncertainty $\sigma_B$ for the $150\,$GHz auto-spectrum, cross-spectrum, and $220\,$GHz auto-spectrum of the SPT field. 
The quoted uncertainties include instrumental noise and the Gaussian sample variance of the primary CMB and the point source foregrounds. 
The sample variance of the SZ effect, beam uncertainty, and calibration uncertainty is not included. 
Beam uncertainties are shown in Figure~\ref{fig:beam} and calibration uncertainties are quoted in \S\ref{sec:calibration}.
Point sources above $6.4\,$mJy at $150\,$GHz have been masked out in this analysis. 
This flux cut substantially reduces the contribution of radio sources to the bandpowers, although DSFGs below this threshold contribute significantly to the bandpowers. }
\normalsize
\end{center}
\end{table*}

\begin{table*}[ht!]

\begin{center}
\caption{\label{tab:bandpowers} DSFG-subtracted Bandpowers}
\small
\label{tab:DSFGsubbandpowers}
\begin{tabular}{cccc}
\hline\hline
\rule[-2mm]{0mm}{6mm}

$\ell$ range&$l_{\rm eff}$ &$q$ ($\mu{\rm K}^2$)& $\sigma$ ($\mu{\rm K}^2$) \\
\hline
2001 - 2200 & 2058 &  221.3 &  16.9  \\ 
2201 - 2400 & 2276 &  130.2 &  11.2  \\ 
2401 - 2600 & 2474 &  126.5 &  10.3  \\ 
2601 - 2800 & 2677 &   60.2 &   7.7  \\ 
2801 - 3000 & 2893 &   50.4 &   8.0  \\ 
3001 - 3400 & 3184 &   36.6 &   5.9  \\ 
3401 - 3800 & 3581 &   21.0 &   6.5  \\ 
3801 - 4200 & 3992 &   22.9 &   8.4  \\ 
4201 - 4600 & 4401 &    8.1 &   9.5  \\ 
4601 - 5000 & 4789 &    3.0 &  11.3  \\ 
5001 - 5900 & 5448 &   11.2 &  10.5  \\ 
5901 - 6800 & 6359 &   16.0 &  16.2  \\ 
6801 - 7700 & 7255 &   60.3 &  27.9  \\ 
7701 - 8600 & 8161 &   32.1 &  42.9  \\ 
8601 - 9500 & 9059 &   54.7 &  63.8  \\ 
\hline
\end{tabular}
\tablecomments{ Band multipole range and weighted value $\ell_{\rm eff}$, bandpower $q_B$, 
and uncertainty $\sigma_B$ for the DSFG-subtracted maps of the SPT field.  
These bandpowers correspond to a linear combination (see \S\ref{subsec:freqsub}) of the 150, $150\times220$, 
and $220\,$GHz power spectra, optimized to remove emission from DSFGs below the point source detection threshold of SPT.  
Point sources above $6.4\,$mJy at $150\,$GHz have been masked out in this analysis.  
The quoted uncertainties include instrumental noise and Gaussian sample variance of the primary CMB and point source foregrounds. 
The sample variance of the SZ effect, beam uncertainty and calibration uncertainty is not included. 
Beam and calibration uncertainties are quoted in \S\ref{sec:beams} and \S\ref{sec:calibration} and shown in Figure~\ref{fig:cldif}.}
\normalsize
\end{center}
\end{table*}

The SPT bandpowers ae dominated by the damping tail of the primary CMB anisotropy on angular multipoles $2000<\ell<3000$. 
At these multipoles, the bandpowers are in excellent agreement with the predictions of a $\Lambda$CDM model determined from CMB observations on larger angular scales.  
On smaller scales, the SPT bandpowers provide new information on secondary CMB anisotropies and foregrounds which 
dominate the primary CMB anisotropy. 
The SPT data presented here represent the first highly significant detection of power at these frequencies and angular scales where the primary CMB anisotropy is sub-dominant. After masking bright point sources, the total signal-to-noise ratios on power in excess of the primary CMB are 55, 55, and 45 at 150, $150\times220$, and 220$\,$GHz respectively. 
The majority of the high-$\ell$ power can be attributed to a Poisson distribution of point sources (likely DSFGs) on the sky. 
H09 use the bandpowers presented in this work to measure both Poisson and clustered contributions to the measured powers
and investigate the implications for the properties of DSFGs. 
The largest source of secondary CMB anisotropy at $150\,$GHz is expected to be the SZ effect, and we investigate 
SZ constraints in the following sections.

\begin{figure*}[ht]\centering
\includegraphics[width=0.9\textwidth]{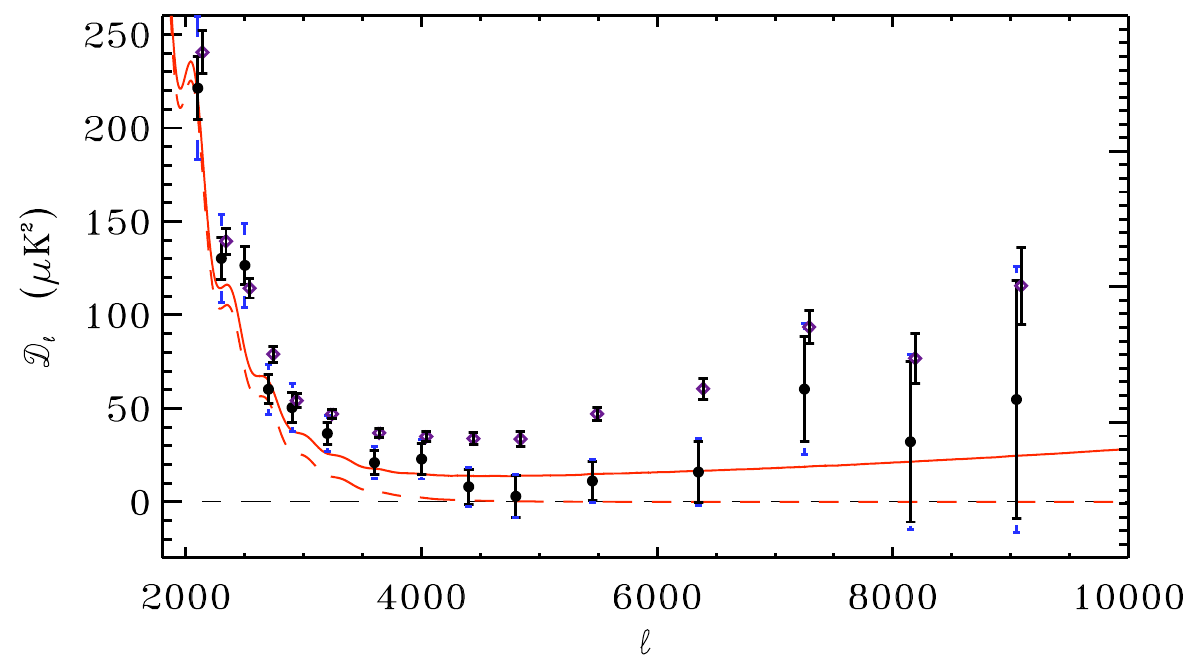}
  \caption[]{The SPT $150\,$GHz ({\bf purple diamonds}) and DSFG-subtracted ({\bf black circles}) bandpowers over-plotted on the best-fit models to the DSFG-subtracted bandpowers.
  The best-fit, lensed $\Lambda$CDM cosmological model for the primary CMB anisotropy is shown by the {\bf dashed red line}, while the sum of the best-fit $\Lambda$CDM model, kSZ, tSZ and point source terms is shown by the {\bf solid red line}. 
  The primary CMB anisotropy alone is a poor fit to the SPT data. 
  The uncertainties on the DSFG-subtracted bandpowers are larger for two reasons. 
  First, the normalization convention inflates the uncertainties by a factor of $1/0.675^2$, and second, these bandpowers
  also include the more noisy $220\,$GHz data.
  Beam and calibration uncertainties are marked by a second blue error bar for the DSFG-subtracted bandpowers only. 
  Note that the calibration and beam uncertainties are correlated between $\ell$-bins. 
  The $150\,$GHz data has been shifted to the right by $\Delta\ell = 40$ for clarity. 
  Point sources above $6.4\,$mJy at $150\,$GHz have been masked in this analysis. }
\label{fig:cldif}
\end{figure*}

\subsection{DSFG-subtracted Bandpowers}\label{sec:ptsrcsub} 

Our immediate goal is to measure the amplitude of the tSZ power spectrum. 
However, several signals in these maps have similar angular power spectrum shapes, and the single-frequency maps only
constrain the sum of the power from these sources.
For instance, the $150\,$GHz data effectively constrain the sum of the tSZ, kSZ, and clustered DSFG power. 

As discussed earlier, each of these components has a distinct frequency dependence so a linear combination of SPT's two frequency bands 
can be constructed (following \S\ref{subsec:freqsub}) to minimize any one of them.  
H09 find significant evidence for a clustered DSFG power contribution to the single-frequency bandpowers listed in Table \ref{tab:bandpowerssinglefreq} with an amplitude comparable to that of the tSZ effect. 
We expect the kSZ effect to be smaller than the tSZ effect at $150\,$GHz on theoretical grounds.
Additionally, due to the frequency dependencies of the components, removing the kSZ effect would inflate the relative contribution of clustered DSFGs with respect to the tSZ effect. 
Therefore, we choose to remove DSFGs from the SPT bandpowers.

For a mean DSFG spectral index, $\alpha$, the proper weighting ratio, $x$, to apply to the $220\,$GHz spectrum for DSFG removal would be: 
\begin{equation}
x =\frac{S_{150}/(\frac{dB_\nu}{dT_{\rm CMB}}|_{150})} {S_{220}/(\frac{dB_\nu}{dT_{\rm CMB}}|_{220})}= (150/ 220)^\alpha \frac{\frac{dB_\nu}{dT_{\rm CMB}}|_{220}}{\frac{dB_\nu}{dT_{\rm CMB}}|_{150}}.
\end{equation}
The spectrum in Table \ref{tab:DSFGsubbandpowers} and Figures \ref{fig:clall} and \ref{fig:cldif} is produced with a weighting factor of $x=0.325$ corresponding to a mean spectral index of $\alpha=3.6$. 
The contribution from DSFGs can be completely removed only if every galaxy has the same spectral index. However, the comparative closeness of SPT's two frequency bands ensures that power leakage into the difference 
maps remains small despite some expected scatter in the spectral index of the dusty galaxies. 
In \S\ref{sec:ptsrcprior}, we motivate this choice of $x$ and discuss predictions for the residual point source power.

We assume that there is no tSZ contribution to the 220$\,$GHz data as the 220$\,$GHz band is designed to be centered on the SZ null. 
 Fourier transform spectroscopy measurements of the 220$\,$GHz band pass confirm that the tSZ amplitude in the 220$\,$GHz band will be $\lesssim\,$5\% of the 150$\,$GHz amplitude.   
 Any error incurred by subtracting roughly one third of the 220$\,$GHz amplitude from the 150$\,$GHz data would be less than $3\%$, far below the present $\sim$$\,40\%$ statistical uncertainty on $A_{\rm SZ}$ (see Table 3). 
 
 It is important to note that the apparent tSZ power in the DSFG-subtracted bandpowers will be a factor of $1/(1-x)^2=2.2$ higher than at 150$\,$GHz as the differenced bandpowers have been normalized to preserve the amplitude of the primary CMB anisotropy.  In this work, we report SZ amplitudes scaled to 153$\,$GHz which is the effective band center of the 150$\,$GHz band for a tSZ spectrum.

\subsubsection{Residual Point Source Power}\label{sec:ptsrcprior} 

The DSFG-subtracted maps have substantially less power due to both unclustered and clustered point sources as seen in Figure~\ref{fig:cldif}. 
However, we include a Poisson point source amplitude in all fits since we expect a small fraction of the point source power to remain in the DSFG-subtracted maps.   
The best-fit amplitude is consistent with zero and unphysical negative values of the Poisson point source power are allowed due to noise. 
To prevent this, we place a prior on the residual Poisson point source power based on what we know about the observed DSFG population from H09 and radio source population from V09 and \citet{dezotti05}.

The residual power in the subtracted map due to Poisson-distributed DSFGs, $C_\ell^{\rm DSFG}$,  depends on the scatter in spectral indices $\sigma_\alpha$, the accuracy to which the mean spectral index $\bar{\alpha}$ is known, and an estimate of the Poisson DSFG power in the $150\,$GHz band, $C_\ell^{{\rm ps},150}$.  
For a given combination of these parameters, this residual DSFG power will be:
\begin{eqnarray} \label{eqn:psprior}
C_\ell^{\rm DSFG} &=& C_\ell^{{\rm ps},150} \times \\
&&\left(\sigma_\alpha^2 \left[ln(\nu_{150}/ \nu_{220})\right]^2 +  \left( 1 - \frac{x}{x_{\rm true}} \right)^2 \right),  \nonumber
\end{eqnarray} 
where $\nu_{150}$ and $\nu_{220}$ are the effective bandcenters of the 150 and 220$\,$GHz bands, and $x$ and $x_{\rm true}$ are the assumed and true values of map weighting ratio.

\begin{figure}[ht]\centering
\includegraphics[width=0.45\textwidth]{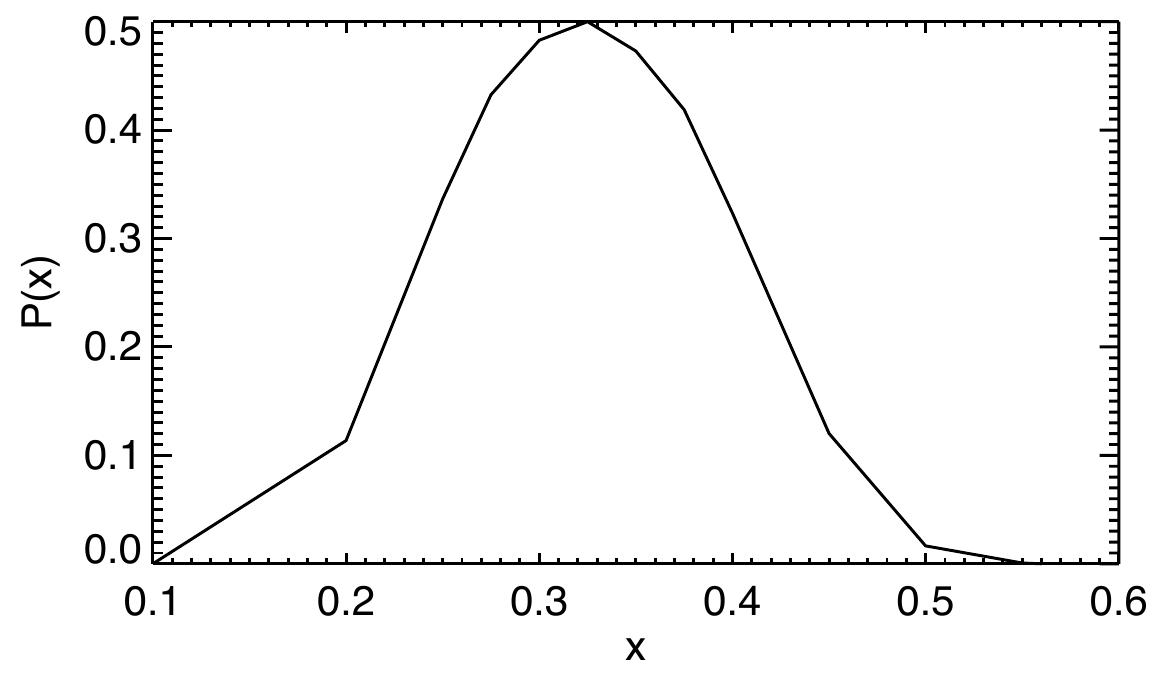}
  \caption[]{Probability that the residual point source power in the DSFG-subtracted map constructed by $\bar{m}_{150} - x \bar{m}_{220}$ is lower than the value at $x = 0.325$ as a function of $x$. We can interpret
this as the probability that a given value of $x$ is the true value.   
  There is a broad maximum centered at 0.325, which corresponds to a spectral index of 3.6 between the 150 and 220$\,$GHz bands. 
  This is consistent with the ratio of point source power between single-frequency fits to the 150, $150\times220$, and 220$\,$GHz bandpowers. 
  We estimate the $x$ which minimizes residual point source power to be $0.325 \pm 0.08$. }
\label{fig:optsubtraction}
\end{figure}

We examine the residual Poisson point source amplitudes for a broad range of weighting ratios to estimate the optimal $x$ value and the error in that estimate. 
For each $x$, we estimate the probability that $C_\ell^{\rm ps}$($x$) is less than $C_\ell^{\rm ps}$($x$ = 0.325) using the MCMC chains described in \S\ref{sec:params}. 
The resulting probability distribution is taken to be the likelihood function for $x_{\rm true}$.
As shown in Figure~\ref{fig:optsubtraction}, there is a broad maximum for $x$ = 0.25 to 0.4 and we adopt the best fit, $x=0.325$, for the following results. 
In the absence of a direct measurement, we place a conservative uniform prior on the scatter in DSFG spectral indices, $0.2<\sigma_\alpha<0.7$, as discussed in H09. 

Our expectation for the residual radio contribution to the DSFG-subtracted bandpowers, $C_\ell^{\rm radio}$, is based on the \citet{dezotti05} radio source count model.  
This model is in excellent agreement  on the high flux end with the SPT source counts (V09).  
As discussed in H09, the residual radio source power after masking is expected to be a small fraction of the DSFG power at $150\,$GHz. 
However, this small radio contribution may be comparable to the residual DSFG power in the DSFG-subtracted spectrum. 
The radio source power can be calculated from the integral of $S^2 dN/dS$ for the \citet{dezotti05} counts model from 
zero to the flux masking threshold of $6.4\,$mJy. 
We compute the power these sources contribute to the optimal DSFG-subtracted spectrum to be $3.9\times10^{-7} \,\mu{\rm K}^2$ by assuming an average spectral index of $\alpha=-0.5$ based on the detected sources in V09.  
This power level is nearly identical to that predicted by the \citet{sehgal09} simulations.  
To allow for a variation in spectral index as well as uncertainty in the model normalization when extended to lower flux sources, we assign a conservative uncertainty of 50\% on the predicted residual radio source power in the prior.

We combine this information to create a prior on the residual point source power in 
the DSFG-subtracted maps $C_\ell^{\rm ps}=C_\ell^{\rm DSFG}+C_\ell^{\rm radio}$.
This prior spans the range $C_\ell^{\rm ps} \in $[3.5,   9.0]$\,\times 10^{-7}\,\mu$K$^2$ at 68\% confidence and [1.2,   13.9]$\,\times 10^{-7}\,\mu$K$^2$ at 95\% confidence.  
The best-fit value of the residual Poisson component before applying the prior is $C_\ell^{\rm ps}$(no prior) = $(6.2 \pm 6.4) \times 10^{-7} \,\mu$K$^2$ and lies at the middle of our assumed prior range.
The upper end of the 95\% range is approximately 20\% of the best-fit value of the Poisson point source power in the undifferenced $150\,$GHz bandpowers. 
This suggests that we have subtracted over 80\% of the point source power from the $150\,$GHz spectrum, with the residual point source power largely from radio sources.  
Without this prior on the Poisson point source amplitude, the uncertainty on the $A_{\rm SZ}$ detection presented in the next section would increase by $\sim 50\%$.

\subsubsection{Residual Clustered Point Source Power}

We assume that the contribution of clustered point sources is insignificant in the DSFG-subtracted bandpowers.
Using a combination of the SPT bandpowers at 150, 220, and $150\times 220\,$GHz 
as well as the DSFG-subtracted spectrum, H09 argue that the residual clustered DSFG component in the DSFG-subtracted bandpowers is 
less than 0.3$\,\mu$K$^2$ at 95\% confidence. 
This is several percent of the SZ power spectrum but negligible at the current detection significance of $\lesssim 3\, \sigma$. 
We also argue in \S\ref{sec:foregrounds} that clustered radio sources are negligible.
Therefore, residual power from clustered point sources will not bias SZ constraints from the DSFG-subtracted bandpowers.

The above argument holds if the point sources are uncorrelated with the SZ signal. 
However, if the clustered term was completely anti-correlated with the SZ signal, the measured SZ power in Table \ref{tab:sigma8} could underestimate the true SZ power by 38\%.  This is unlikely for two reasons. 
First, the residual after DSFG subtraction should be uncorrelated as long as the spectral dependence of 
cluster member DSFGs is similar to the general DSFG population.  
Second, as argued in \S\ref{sec:foregrounds}, most DSFGs are not galaxy cluster members. 
We also look at the correlations between the \citet{sehgal09} DSFG and tSZ simulated sky maps.
The \citet{sehgal09} DSFG model scales the number density of DSFG cluster members linearly with cluster mass; this is a substantially stronger scaling than observed \citep{bai07}.
Therefore, estimating the cluster-DSFG correlation from the \citet{sehgal09} should be overly conservative.
We calculate the anti-correlation coefficient between the \citet{sehgal09} simulated tSZ maps at $148\,$GHz and a linear combination of the 148 and $219\,$GHz simulated IR source maps with the same weighting as used for the SPT DSFG-subtracted bandpowers.
We find the anti-correlation coefficient between the tSZ effect and total DSFG power to be 21\% using the power measured at $\ell = 3000$. 
This highly conservative upper limit of 21\% on the anti-correlation implies that the true SZ power is underestimated by less than 4\%.
Based on these arguments, we assume that correlations between SZ signals and emission from cluster member galaxies is negligible in this analysis.

Separately, one might worry about correlations between radio sources and SZ clusters. 
Radio sources could suppress the tSZ signal by ``filling in'' the tSZ decrements. 
However, the work by \citet{lin09} and \citet{sehgal09} shows the number of cluster-correlated radio sources is expected to be small. 
For instance, we can examine this correlation in the simulated tSZ and radio source maps produced by \citet{sehgal09}.  
We look at the power spectrum at $\ell = 3000$ for the tSZ map, the tSZ+radio-source map, and the radio source only map after masking sources above $6.4\,$mJy.  
We find an anti-correlation coefficient of 2.3\% for  the two components. 
Given the expected radio source power level of $\sim\,0.6\,\mu{\rm K}^2$ at $\ell = 3000$ (\S\ref{sec:ptsrcprior}), radio source-cluster correlations should not affect the results in this work.

\subsection{Markov Chain Analysis}
\label{sec:params}

The DSFG-subtracted bandpowers presented in \S\ref{sec:ptsrcsub} detect at high significance 
a combination of the primary CMB anisotropy, secondary SZ anisotropies, and residual 
point sources.
In this section, we use an MCMC analysis to separate these three components and to produce an unbiased measurement of the tSZ power spectrum amplitude. 

\subsubsection{Elements of the MCMC Analysis}
\label{subsec:mcmc}

We fit the DSFG-subtracted bandpowers to a model including the lensed primary CMB anisotropy, secondary tSZ and kSZ anisotropies, and a residual Poisson point source term.
We use the standard, six-parameter, spatially flat, lensed 
$\Lambda$CDM cosmological model to predict the primary CMB temperature anisotropy. 
The six parameters are the baryon 
density $\Omega_b$, the density of cold dark matter $\Omega_c$, the optical 
depth to recombination $\tau$, the angular scale of the peaks $\Theta$, the 
amplitude of the primordial density fluctuations $\ln[10^{10} A_s]$, and the 
scalar spectral index $n_s$.    
To fit the high-$\ell$ power, we extend the basic, six-parameter model with two additional parameters: the amplitude of a tSZ power spectrum template, $A_{\rm SZ}$, and a constant, $C_\ell^{\rm ps}$, such as  would be produced by a Poisson distribution of point sources on the sky.   
We also explore the potential impact of the kSZ effect on these parameters by using three different kSZ models.

Gravitational lensing of CMB anisotropy 
by large scale structure tends to increase the power at small angular scales, with the
potential to influence a SZ power spectrum measurement.  The
calculation of lensed CMB spectra out to $\ell=10000$ proved
prohibitively expensive in computational time.  We avoid this
computational limitation by calculating the lensing contribution for
the the best-fit cosmological model, and adding this estimated
lensing contribution to the unlensed CMB power spectrum calculated at each step in the Markov chain. 
Given the small allowed range of $\Omega_m$ with current CMB data,
we predict that using the fixed lensing contribution will misestimate the actual lensing by less than 30\%.  
We have checked this assertion on a sampling of parameter sets drawn from the chains. 
The lensing contribution to the high-$\ell$ spectrum is $\sim$1.5 $\mu{\rm K}^2$ while the modeled tSZ spectrum for the same best-fit WMAP5 cosmology averages $\sim\,8.6\,\mu{\rm K}^2$ near $\ell = 3000$ where SPT has the highest S/N on the SZ spectrum.
 In the differenced spectra used to derive the $A_{\rm SZ}$ constraints (see
\S\ref{sec:ptsrcsub}), the ratio of lensing to the tSZ effect is
suppressed by a factor of $(1 - x )^2 = 0.46$.  
This reduction occurs because the subtracted spectrum is normalized such that the CMB power is unaffected, though the SZ power is enhanced.  
As a result, we expect possible lensing misestimates to introduce a negligible error of $\lesssim\,$3\% on the $A_{\rm SZ}$ constraints. Of course, this error will be larger for smaller values of $A_{\rm SZ}$.

The tSZ template we use as a fiducial model is based on simulations by \citet{sehgal09}.\footnote{http://lambda.gsfc.nasa.gov/toolbox/tb\_cmbsim\_ov.cfm}
The simulations are for a WMAP5 cosmology with $\sigma_8 = 0.80$ and $\Omega_b h = 0.0312$ and an 
observing frequency of $148\,$GHz. 
We rescale the template to $153\,$GHz, the effective center frequency of the SPT 150$\,$GHz band for signals
with a tSZ spectrum.

For a limited number of
cases, we also compare the results of the Sehgal numerical tSZ template to those of 
the Komatsu \& Seljak analytic template with WMAP5 parameters\footnote{http://lambda.gsfc.nasa.gov/product/map/dr3/pow\_sz\_spec\_get.cfm}
\citep{komatsu02} and the numerical template by \citet{shaw09}.
The primary difference between the Shaw and Sehgal simulations is the value of the energy feedback parameter in the \citet{bode07} intracluster gas model. 
The Shaw simulations have the feedback parameter reduced to 60\% of the Sehgal value. 
Increasing the feedback parameter causes the gas distribution in clusters to `puff-out' or inflate; in low mass clusters the gas may become unbound altogether. The overall effect is to reduce the  predicted tSZ power spectrum, especially at small angular scales. 
We limit our analysis of specific tSZ models to these three models.

An independent measurement of the kSZ spectrum is outside the scope of this work, however,
it is necessary to take the kSZ effect into account when inferring the tSZ amplitude from the data. 
We consider three kSZ cases based on two published kSZ models.  
The kSZ effect is assumed to be zero in the first case (``no kSZ").
As a second, intermediate case, we use the model by \citet{sehgal09} (referred to as ``homogeneous kSZ").   
This model includes kSZ contributions from a homogeneous reionization scenario, but does not include the additional kSZ power produced by patchy reionization scenarios.   We take this case to be the fiducial kSZ model.
We include an estimate of patchy reionization in the third kSZ case.
For the patchy reionization phase, we use the ``brief history" model B from \citet{zahn05}, re-calculated for WMAP-5 best fit cosmological parameters. 
The sum of the homogeneous kSZ model and the patchy contribution will be referred to as the 
``patchy kSZ" model.  
We do not scale these templates for different cosmological parameter sets as we expect the kSZ theoretical uncertainty to be at least as large as the cosmological dependence.
Finally, we include a residual Poisson point source component in all chains. 
The construction of the point source prior is outlined in \S\ref{sec:ptsrcprior}. 

Previous CMB experiments have produced exceptional constraints on the primary CMB anisotropy, and we use the bandpowers from WMAP5 \citep{dunkley09}, ACBAR \citep{reichardt09a}, and QUaD \citep{brown09} at $\ell < 2200$ in all parameter fitting. 
We refer to this collection along with the SPT DSFG-subtracted bandpowers as the `CMBall' data set. 
It is important to note that the two-parameter extension to the $\Lambda$CDM model for point sources and the tSZ effect is only restricted to the SPT bandpowers. 
This restriction is imposed for two reasons.
First, the point source contributions to each experiment may be different due to the different frequencies and flux cuts for masking
sources. 
Arguably, the point source power is likely to be similar
in the 150$\,$GHz SPT, ACBAR, and QUaD results, but we would be unable to use frequency information to
discriminate between the SZ effect and clustered DSFGs. 
Second, the primary CMB is dominant below $\ell \sim 3000$ and the other experiments lack sufficient statistical weight at high-$\ell$ to
improve upon SPT's measurement of SZ effect and point source power.

Parameter estimation is performed by MCMC sampling of the full multi-dimensional parameter space using an extension of the {\textsc CosmoMC} package \citep{lewis02b}. 
We include the code extension produced by the QUaD collaboration \citep{brown09} to handle uncertainties on a non-Gaussian beam in {\textsc CosmoMC}. 
CMB power spectra for a given parameter set are calculated with {\textsc CAMB} \citep{lewis00}.  
We use the WMAP5 likelihood code publicly available from http://lambda.gsfc.nasa.gov.   
After the burn-in period, each set of four chains is run until the largest eigenvalue of the Gelman-Rubin test is smaller than 0.0005. 
Wide uniform priors are used on all six parameters of the $\Lambda$CDM model.  
A weak prior on the age of the Universe ($t_0 \in [10, 20]$ Gyrs) and Hubble constant ($h \in [0.4, 1]$) is included in all chains, but should not affect the results. 
We use a uniform prior on $A_{\rm SZ}$ over a wide range from -1 to 10 times the value expected for $\sigma_8 = 0.80$.

\subsubsection{Constraints on SZ amplitude}

We fit for the normalization factor of a fixed tSZ template and, in Table \ref{tab:sigma8},
report both this template-specific normalization, $A_{\rm SZ}$, and the total inferred SZ-power at $\ell = 3000$, 
near the multipole with maximum tSZ detection significance. 
This estimate of the SZ power includes both tSZ and kSZ terms, and is included to facilitate comparison with other SZ models.
We expect both the thermal and the kinetic SZ spectra to vary slowly with angular multipole.

\begin{table*}[ht!]
\begin{center}
\caption{\label{tab:sigma8} Constraints on $A_{\rm SZ}$ and $\sigma_8$}
\small
\begin{tabular}{lcc}
\hline\hline
\rule[-2mm]{0mm}{6mm}
& primary CMB & CMB + $A_{\rm SZ}$:\\
\hline
$A_{\rm SZ}$: & - & $0.55 \pm 0.21$ \\
$A_{\rm SZ}$  (w homogeneous kSZ):& - & $0.42 \pm 0.21$ \\
$A_{\rm SZ}$  (w patchy kSZ):& - & $0.34 \pm 0.21$ \\
\hline
SZ power at $\ell = 3000$ &  \\
( tSZ + 0.46 $\times$ kSZ): & - &$4.2 \pm 1.5\, \mu{\rm K}^2$\\
\hline
kSZ power at $\ell = 3000$ &  \\
 ~~~homogeneous kSZ && $2.0\, \mu{\rm K}^2$ \\
 ~~~patchy kSZ &&  $3.3\, \mu{\rm K}^2$\\
\hline
$\sigma_8$ (no kSZ): & $0.795 \pm 0.033$ &   $0.778 \pm 0.024$\\
$\sigma_8$ (w homogeneous kSZ): &$0.794 \pm 0.028$   & $0.773 \pm 0.025$ \\
$\sigma_8$ (w patchy kSZ): & $0.788 \pm 0.029$  &  $0.770 \pm 0.024$\\
\hline\hline
\end{tabular}
\vspace{1em}
\tablecomments{ The $1\,\sigma$ constraints on $\sigma_8$ derived from the DSFG-subtracted analysis of the SPT data, when using the simulations in \citet{shaw09} to estimate the non-Gaussian cosmic variance of the tSZ power spectrum.  
The best-fit value for the amplitude of the tSZ power spectrum is also shown, normalized to unity for a WMAP5 cosmology with $\sigma_8 = 0.8$.
$A_{\rm SZ} = 1$ corresponds to a power of $7.5\,\mu{\rm K}^2$ at $\ell=3000$.
Results are shown with cosmic variance added in quadrature to the statistical uncertainty, however, the A$_{\rm SZ}$ constraints are dominated by statistical uncertainties.
Results are shown for no kSZ effect, for a homogeneous model of the kSZ effect \citep{sehgal09} and the homogeneous model with an additional patchy reionization power contribution \citep{zahn05}.  
Finally, the joint constraint on the combined kSZ/tSZ power is shown under the assumption that the two templates are effectively degenerate.
For reference, we also quote the power of the two kSZ models considered.
The SPT data constrains the combined amplitude of the SZ contributions, which we quote at $\ell = 3000$ where 
the measurements have the most constraining power.
The kSZ and tSZ receive different pre-factors in the frequency-differenced analysis due to their relative spectral dependence.}
\normalsize
\end{center}
\end{table*}

\begin{figure*}[ht]\centering
\includegraphics[width=0.9\textwidth]{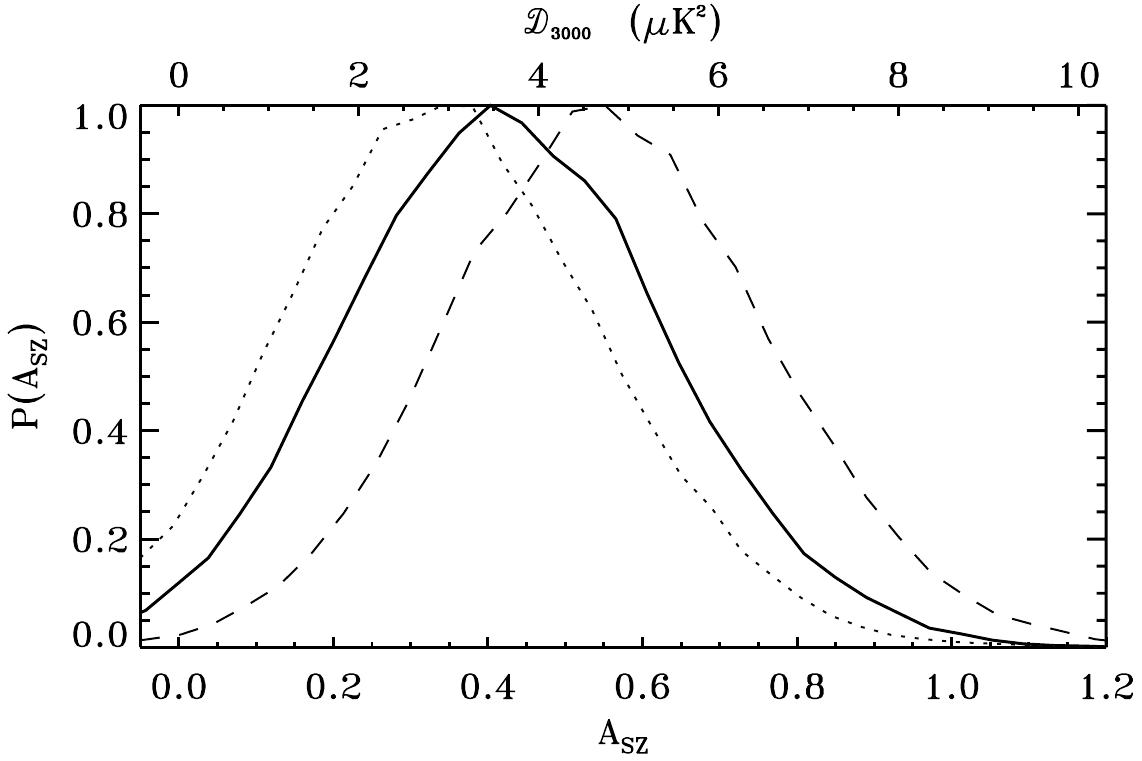}
  \caption[]{The 1D marginalized A$_{\rm SZ}$ constraints from the SPT DSFG-subtracted bandpowers. 
  Three kSZ cases are considered: no kSZ effect ({\bf dashed line}), the homogeneous kSZ model ({\bf solid line}) and the homogeneous model plus a patchy reionization term ({\bf dotted line}). 
  These models are described more fully in \S\ref{subsec:mcmc}. 
    {\em Top axis}: The corresponding tSZ power at $\ell = 3000$ for reference.  The no-kSZ curve (dashed line) can be interpreted as a constraint on the sum of $D_{3000}^{\rm tSZ}$ + 0.46 $\times$ $D_\ell^{\rm kSZ}$.
}
\label{fig:like1d_asz}
\end{figure*}

The chains are run for three different assumptions about the kSZ effect: the no kSZ, homogeneous kSZ, and patchy kSZ models described above. 
In each case, we produce MCMC chains with a fixed kSZ amplitude. 
The $\chi^2$ of the 15 SPT bandpowers is between 16.3 and 16.4 for the best fits of the three kSZ cases considered; there is essentially no impact on the quality of the fit. 
The $A_{\rm SZ}$ and power constraints for each case are listed in Table \ref{tab:sigma8} and plotted in Figure~\ref{fig:like1d_asz}.

\begin{figure*}[ht]\centering
\includegraphics[width=0.9\textwidth]{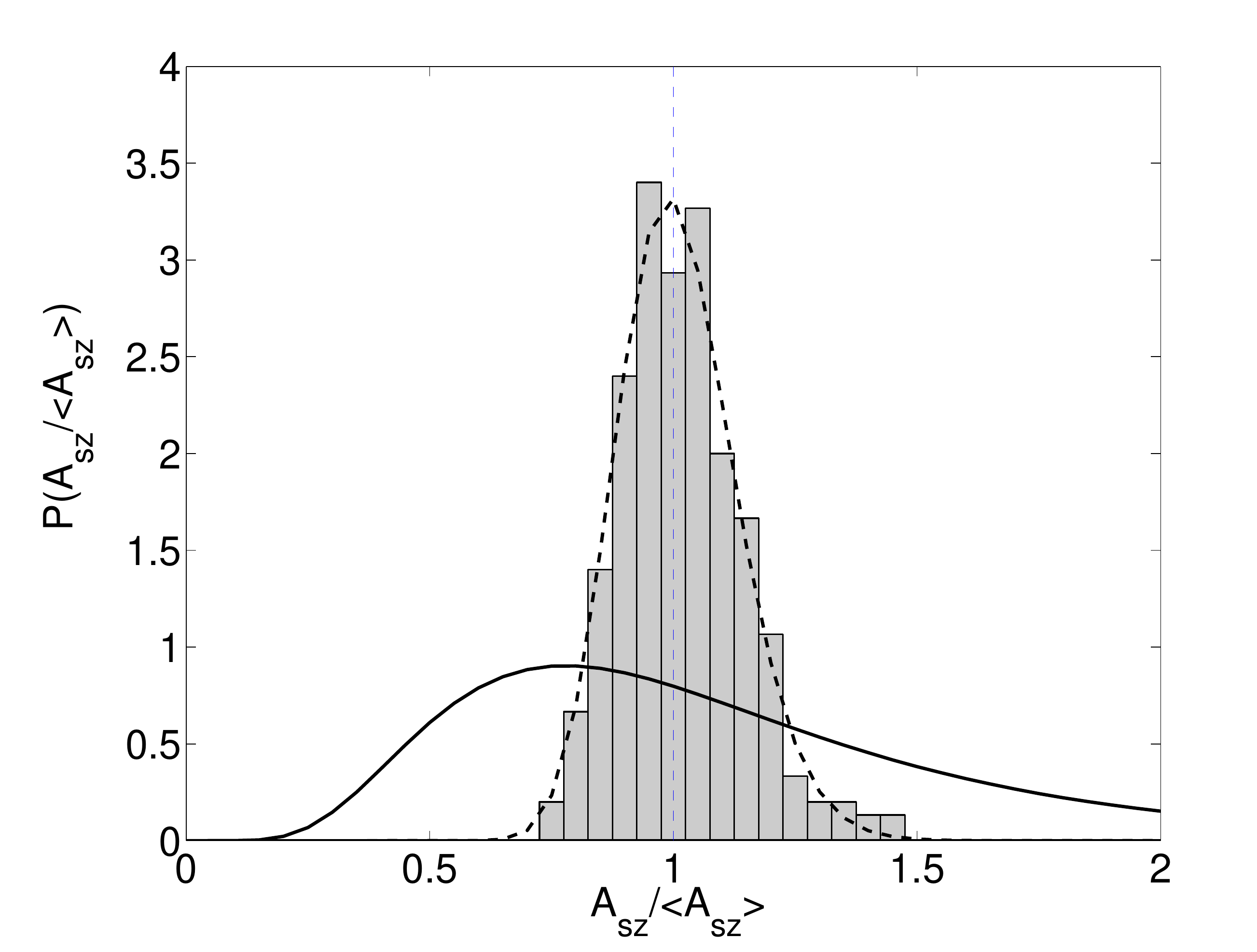}
  \caption[]{Sample variance and assumed theoretical uncertainty on
    the tSZ amplitude.  The histogram shows the sample variance of
    $A_{\rm SZ}/\langle A_{\rm SZ}\rangle$, where $\langle A_{\rm SZ}\rangle$ is
    the mean value measured over a sample of 300 simulated
    maps. The overlying {\bf dashed black line} shows the lognormal
    fit to the distribution, i.e., a Gaussian fit to $\ln (A_{\rm SZ})$),
    with $\sigma_{\ln A} = 0.12$. The {\bf solid black line} is the
    assumed 50\% theoretical uncertainty on $A_{\rm SZ}$, which we model
    as a Gaussian distribution in $\ln (A_{\rm SZ})$.}
\label{fig:szng}
\end{figure*}

The bandpower uncertainties in Table \ref{tab:DSFGsubbandpowers} do not include the sample variance of the tSZ effect, and we must convolve the $A_{\rm SZ}$ distribution in the chains with an estimate of the sample variance to find the true $A_{\rm SZ}$ likelihood function.
The sample variance of the tSZ effect is estimated using the simulations of \citet{shaw09} rerun with the same intracluster gas model parameters as \citet{sehgal09}.
The simulation consists of 300 map realizations of
the size of the SPT sky patch. These are constructed from a base
sample of 40 independent maps by separating the components of each map
into eight redshift bins (between $0 \leq z \leq 3$) and shuffling these bins between
maps to generate a larger sample. We take the spectrum of each
realization and find the best-fit $A_{\rm SZ}$ amplitude after weighting
$\ell$-bins based on the SPT bandpower uncertainties. We use the
distribution of these amplitudes to map out the likelihood function
for the tSZ power (see Figure~\ref{fig:szng}).  Due to the number of independent realizations, we
have limited ability to resolve the tail of the likelihood function.
Fortunately, the non-Gaussianity is small for such a large sky area and
the distribution of $\ln (A_{\rm SZ})$ is well-fit by a Gaussian
with a 12\% width. We use this Gaussian fit as an estimate of the full
likelihood surface.  Small deviations from the true description of the
tSZ sample variance will not impact the final results as the sample
variance is small compared to both the statistical uncertainties and the assumed 50\% model uncertainty. The uncertainties on $A_{\rm SZ}$ are essentially unchanged by the inclusion of the tSZ sample variance.
 The sample variance and model uncertainty are shown in Figure~\ref{fig:szng}.

As discussed earlier, the DSFG-subtracted bandpowers are sensitive to a linear combination of the tSZ and kSZ effects.
We expect analysis of 2009 and later SPT data which include 95$\,$GHz data to be able to separate the two SZ effects. 
The linear combination is not a simple sum, as the frequency-differencing used
to produce the DSFG-subtracted spectrum suppresses the kSZ relative to the tSZ by a factor of  
$(1-x)^2 = 0.46$. 
This factor is uncertain at the 15\% level due to the relative calibration uncertainty between the bands. 
The SPT data detect the combined SZ effect at $2.6\,\sigma$ with tSZ + 0.46 $\times$ kSZ = $4.2\,\pm\,1.5\ \mu{\rm K}^2$ at $\ell = 3000$. 
Using this combined constraint implicitly assumes that the tSZ and kSZ templates are perfectly degenerate, which is a good assumption for the current data quality and templates used in this work.

We can compare the power detected with SPT to that reported by the CBI collaboration \citep{sievers09}. We use the best-fit normalization of a Komatsu \& Seljak template for WMAP5 parameters to compare directly the results of the two experiments. 
There are sub-percent differences in the assumed values of $\sigma_8$ and $\Omega_b h$ 
between the template we adopted here and that used in \citet{sievers09},
which would change the amplitude of the template by $\sim$1\% for an assumed scaling of $\sigma_8^7 (\Omega_b h)^2$. 
This effect is negligible. 
We find the best-fit normalization of the WMAP5 Komatsu \& Seljak model to be $0.37 \pm 0.17$ for the SPT data under the homogeneous kSZ scenario. 
This is $2.4\,\sigma$ below the best-fit CBI normalization of $3.5 \pm 1.3$. 
The model includes the frequency dependence of the tSZ effect. 
The smaller SPT bandpowers suggest that the CBI excess power may be produced by foregrounds with a
frequency dependence falling more steeply than the SZ effect such as radio sources.

\subsection{Implications of the A$_{\rm SZ}$ Measurement} 
\label{sec:sigma8}

The best-fit normalization for the fiducial tSZ spectrum, $A_{\rm SZ}$, is 
significantly lower than unity.  The cosmological parameters assumed when generating this 
template may be slightly different than the best-fit models, so we scale the template to match the 
cosmologies explored by the Markov chain. When these scalings are taken into account, the measured 
values of $A_{\rm SZ}$ are still low.  The tSZ power spectrum depends on the details of how the baryon 
 intracluster gas populates dark matter halos and on cosmology through the number density of these 
 halos.   The paucity of tSZ power may reflect an overestimate of the intracluster gas pressure by the 
 fiducial model and so we compare the template to other SZ models.  At the same time, this low 
 value of $A_{\rm SZ}$ favors a shift in the derived cosmological parameters, particularly $\sigma_8$.

Even a $\sim$$\,2.5\,\sigma$ detection of $A_{\rm SZ}$ will produce cosmologically interesting constraints on $\sigma_8$, as we expect $A_{\rm SZ}$ to scale strongly
with $\sigma_8$ and less strongly with the baryon density. 
The scaling is
approximately $\sigma_8^\gamma (\Omega_b h)^2$ where $7 < \gamma < 9$, depending on
the exact cosmology \citep{komatsu99a, komatsu02}. 
We explore the $\sigma_8$ dependence under the
Press-Schechter halo model and find that this relationship steepens
with the currently favored lower values of $\sigma_8$.  
Sampling
cosmological parameter values from the WMAP5 MCMC chains (which properly treats degeneracies between $\sigma_8$ and other parameters), we find the modeled amplitude of
the tSZ power spectrum varies approximately as $\sigma_8^{11}$.  
Constraining the
amplitude of the tSZ effect offers an independent measurement of $\sigma_8$ that can be compared to measurements based on primary CMB anisotropy or large scale structure. 
Such comparisons test our understanding of the physical processes involved in structure formation.
  
For each point in the MCMC chain, we calculate the predicted $A_{\rm SZ}$ value from the six basic cosmological parameters. 
For this calculation, we use the mass function of \citet{jenkins01} to determine the abundance of galaxy clusters of a given mass. 
We then use the mass-concentration relation of \citet{duffy08} to determine the dark matter halo properties,  
and the gas model used in \citet{komatsu02} to estimate the tSZ signal for each halo according to its mass. 
In order to convert this analytic tSZ spectrum into an amplitude, we take an $\ell$-weighted average designed to match the relative weights each multipole receives in the real tSZ fits.  
This amplitude is normalized to unity for the cosmological parameters assumed in the fiducial tSZ model.
We also allowed the mass-concentration index to vary with cosmology by appropriate scaling of the characteristic mass $M_*$, 
but this was found to be a negligible effect within the explored range in parameters. 

\begin{figure*}[ht]\centering
\plottwo{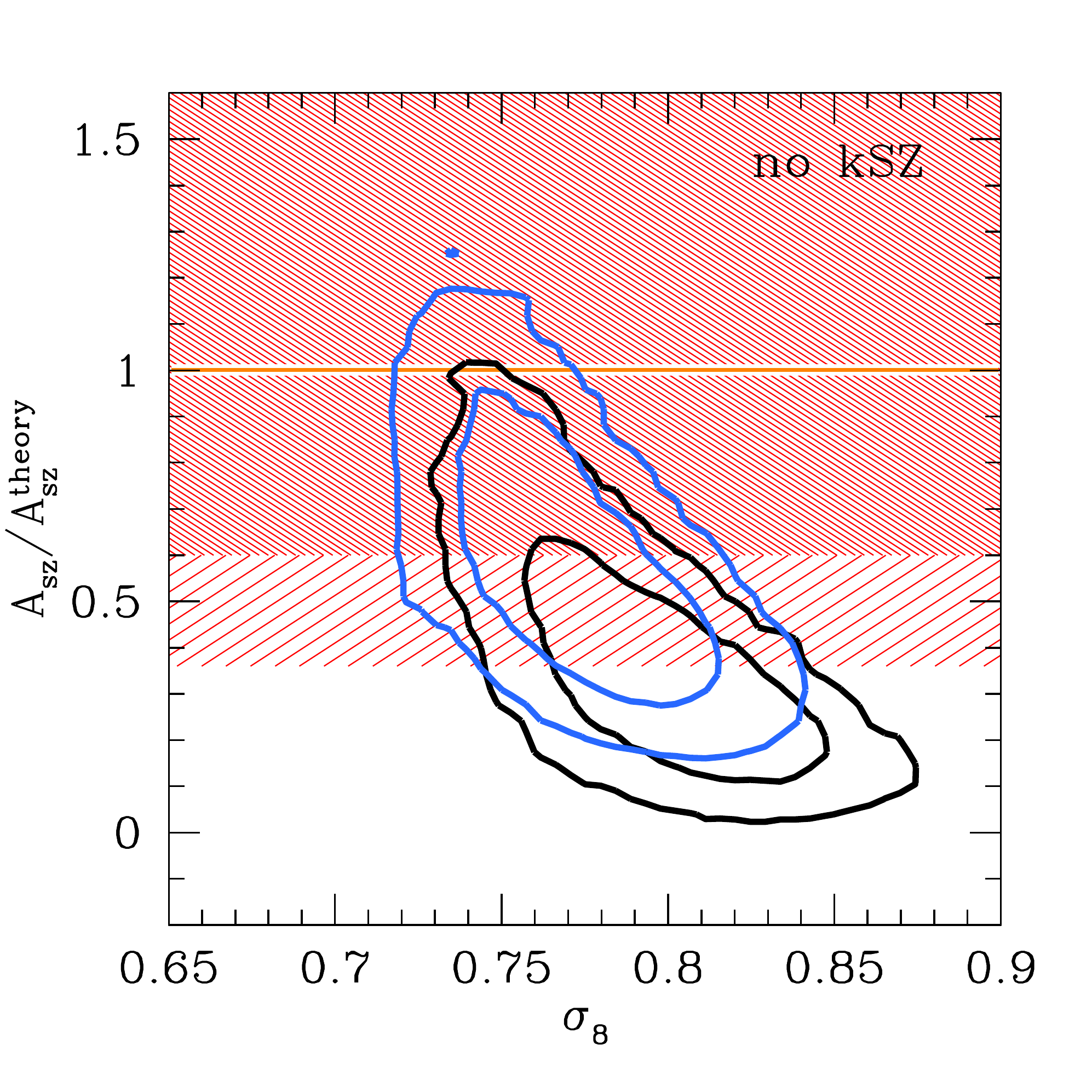}{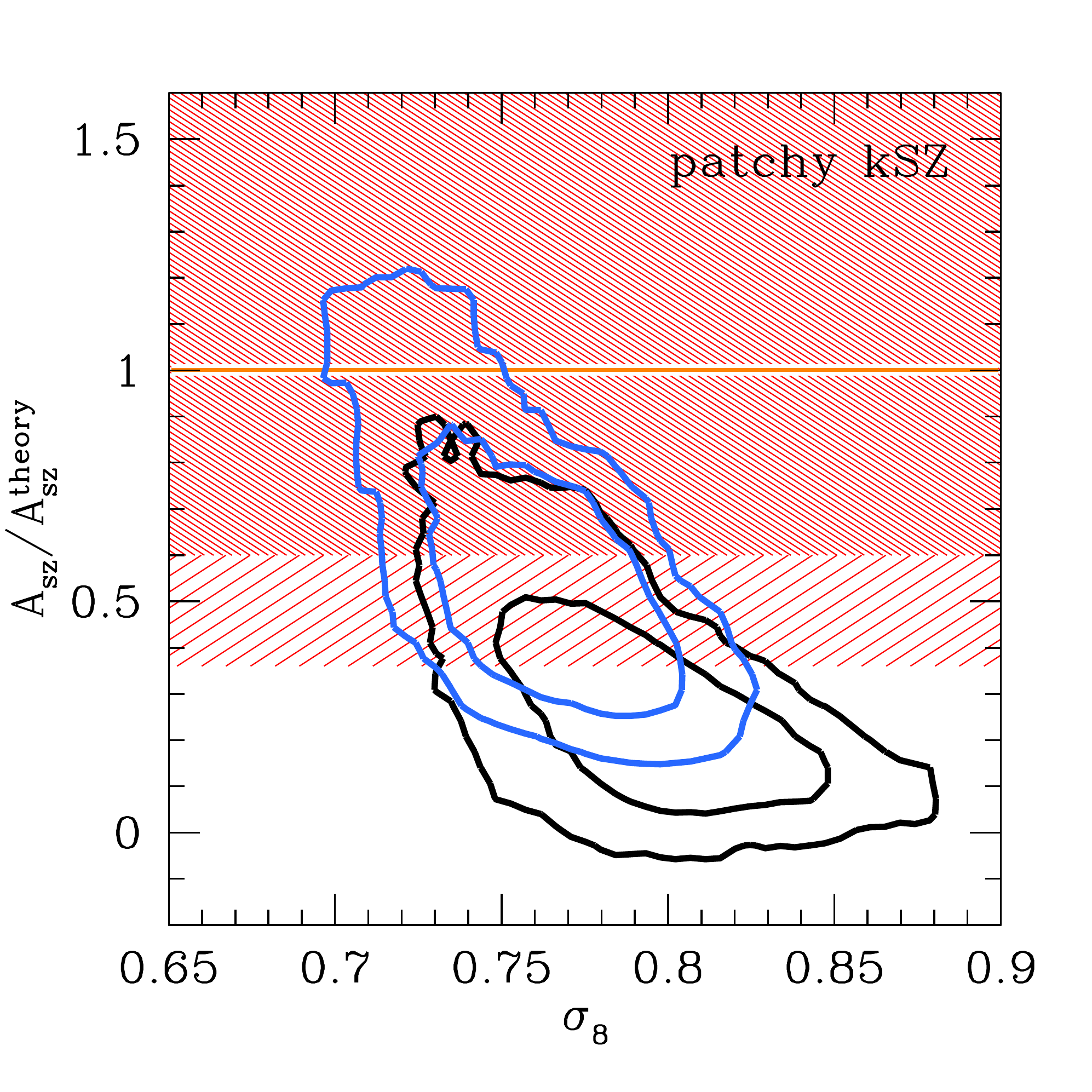}
 \caption[]{ 
Two-dimensional likelihood contours at 68\% and 95\% confidence for $\sigma_8$ versus the tSZ scaling factor, $A_{\rm SZ}$/ $A^{\rm theory}_{\rm SZ}$, derived from the SPT DSFG-subtracted bandpowers.
For each point in the Markov chain, the tSZ scaling factor compares the $A_{\rm SZ}$ value fit to the SPT data to the $A^{\rm theory}_{\rm SZ}$ value predicted for that point's $\Lambda$CDM model parameters (see \S\ref{sec:sigma8}).   
The {\bf black contours} show the likelihood surface for the CMBall dataset.
We observe no dependence between $A_{SZ}$ and the six parameters of the  $\Lambda$CDM model.
The tilt towards higher scaling factors at lower $\sigma_8$ is expected since the predicted $A^{\rm theory}_{\rm SZ}$ depends 
steeply on the value of $\sigma_8$.
The black contours also do not account for the cosmic variance of $A_{\rm SZ}$; without cosmic variance or uncertainty in 
modeling the tSZ power, the tSZ scaling factor would be constrained to be exactly unity ({\bf solid orange line}).
The {\bf red shaded regions} about unity illustrate the uncertainty we assume for the tSZ scaling factor due to theoretical uncertainty 
and cosmic variance (also see Fig. \ref{fig:szng}). 
This uncertainty is modeled as a log-normal distribution.
The measured value of the tSZ scaling relation, including the theoretical uncertainty and sample variance in the model, 
is used to importance sample the Markov chain and obtain the likelihood surface marked by the {\bf blue contours}.
\emph{Left panel:}  Likelihood surfaces assuming no kSZ contribution.  
\emph{Right panel}: Likelihood surfaces assuming the patchy kSZ model.
The constraints for the homogeneous kSZ model will lie between the results for these two cases.  
}
\label{fig:szdeplete}
\end{figure*}

At each point in the chain, the measured tSZ amplitude is compared to the
predicted tSZ amplitude to construct a tSZ scaling factor, ${\rm A}_{\rm SZ}/{\rm A}_{\rm SZ}^{\rm theory}$. 
This procedure will account for any correlations between the measured A$_{\rm SZ}$ parameter and the six $\Lambda$CDM parameters in a self-consistent fashion,
 although we do not see evidence for such correlations in the current data.
The distribution of scaling factors vs.\ $\sigma_8$ is illustrated by the black contours in Figure \ref{fig:szdeplete}.   

In general, the tSZ scaling factors are less than unity.  
These low scaling factors suggest either an over-estimate of the tSZ effect or lower values of $\sigma_8$.
Models predicting larger kSZ or tSZ effects lead to lower scaling factors. 
However, the results can not be purely explained by an over-estimate of the kSZ effect since this tension 
persists in the no-kSZ case. 
Alternatively, the scaling factors may indicate that the explored range in cosmological parameters 
is systematically overestimating the RMS of the mass distribution. 
For instance, as we see in Figure \ref{fig:szdeplete},  points of the chain with lower values of $\sigma_8$, have scaling factors closer to unity. 

The first interpretation of the low tSZ scaling factor is that the Sehgal tSZ template overestimates the tSZ power spectrum.
There is currently some degree of uncertainty in the expected shape
and amplitude of the tSZ power spectrum as predicted by
analytic models or hydrodynamical simulations. 
One reason for this is that cosmological simulations of the intracluster medium have only recently begun to investigate in detail the impact of radiative cooling, non-gravitational heating
sources (such as AGN), and possible regulatory mechanisms between them.
The computational
expense of running hydrodynamical simulations with sufficient
resolution to resolve small-scale processes (such as star-formation)
while encompassing a large enough volume to adequately
sample the halo mass function is prohibitive to a detailed
analysis of the predicted SZ power spectrum. In order to accurately predict the tSZ power spectrum, it is especially important to correctly model the gas temperature and density distribution in low mass ($M < 2\times 10^{14} h^{-1} M_{\odot}$) and high redshift ($z > 1$) clusters, which contribute significantly to the power spectrum at the angular scales where SPT is most sensitive \citep{komatsu02}.

\begin{figure*}[ht]\centering
\includegraphics[width=0.9\textwidth]{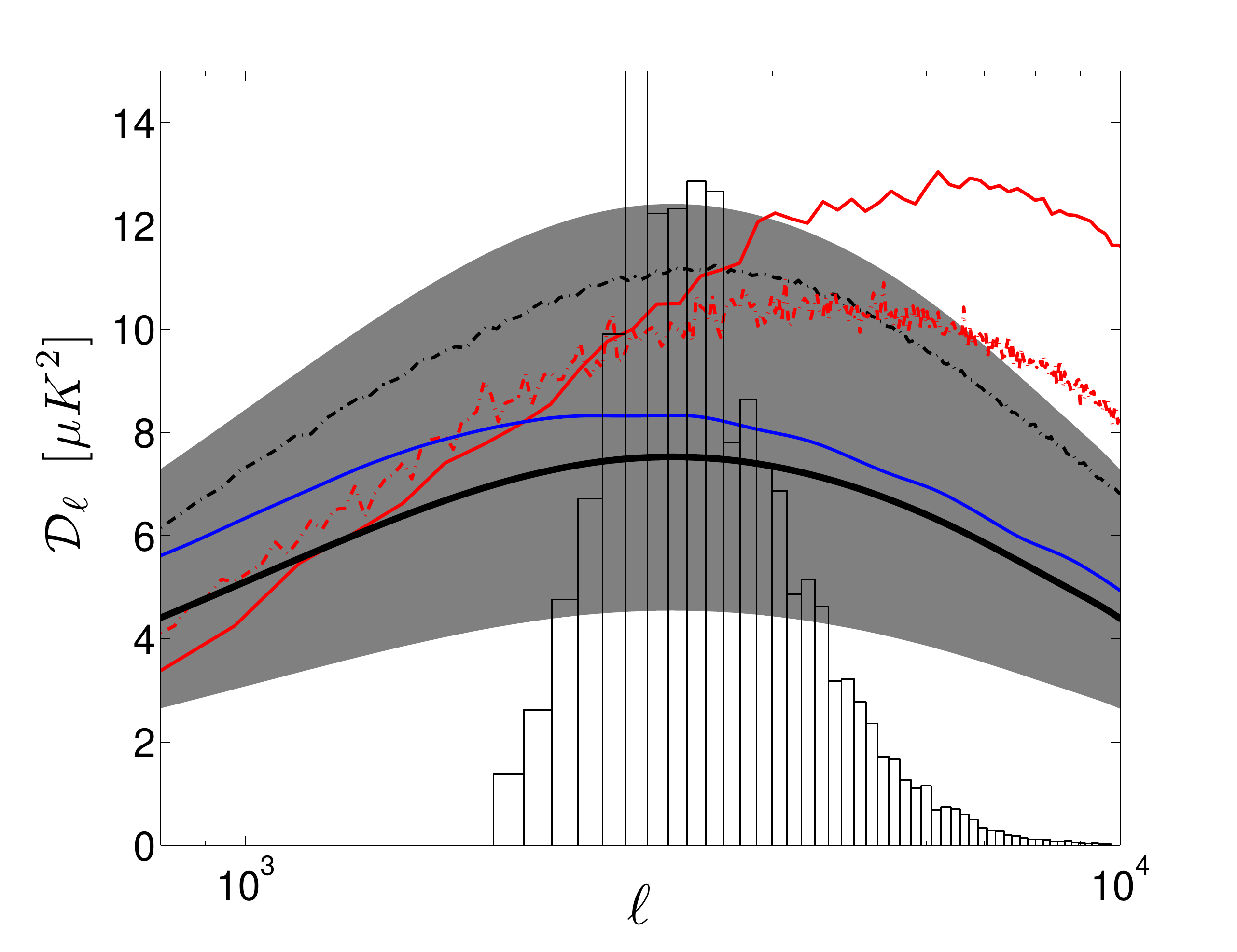}
 \caption[]{
 Comparison of the tSZ power spectrum (at $153\,$GHz)
   as  predicted by numerical simulations and halo model
    calculations. Note that all curves have been re-normalized to the
    fiducial cosmology. 
    The {\bf thick, black, solid line} shows the base template, obtained from maps generated by \citet{sehgal09}.
    The {\bf black, dot-dashed line} shows the modeled spectrum obtained from maps generated
by applying the semi-analytic model for intracluster gas of
\citet{bode07} to the halos identified in an N-body lightcone
simulation \citep[as described in ][]{shaw09}. 
  The {\bf red, solid line} shows the results
    from maps constructed from an adiabatic simulation produced using the Eulerian
    hydrodynamical code CART \citep{kravtsov05}, while the {\bf red,
    dot-dashed line} shows results from maps made from the MareNostrum simulations \citep{zahn10}.
  The {\bf blue, solid line} shows the predictions of the  \citet{komatsu02} halo model calculation. 
  The histogram shows the SPT sensitivity to the thermal SZ signal in each $\ell$-band (with arbitrary normalization) 
The {\bf grey band} illustrates the 68\% confidence interval of our theoretical uncertainty (see Figure \ref{fig:szng}). 
}
\label{fig:theoryprior}
\end{figure*}

In Figure~\ref{fig:theoryprior}, we plot the tSZ power spectrum derived from several different simulations (note that all
curves have been normalized to the fiducial cosmology). 
The thick black solid line shows the base template, obtained from maps generated by \citet{sehgal09}.
The black dot-dashed line shows the power spectrum obtained from the simulations of \cite{shaw09}, which have a lower energy feedback parameter than the \citet{sehgal09} simulations. 
Increasing the amount of feedback energy has
the effect of inflating the gas distribution and
suppressing power at small angular scales. 
The red solid and dot-dashed lines show the power
spectrum from maps constructed from a $240 h^{-1}$ Mpc box simulation run using
the Eulerian hydrodynamics code CART \citep[][Douglas Rudd, private
  communication]{kravtsov05} and from the MareNostrum simulation \citep{zahn10} respectively. 
  Both of these simulations were run in the adiabatic regime, i.e., no cooling, star-formation or feedback. 
The gas distribution is more centrally concentrated than in the Shaw or Sehgal models,  producing tSZ power spectra with significantly more power at small angular scales and less at larger angular scales. 
  The blue solid line shows the analytic template predicted by the \citet{komatsu02} halo model calculation. 
The histogram shows the SPT sensitivity to the thermal SZ signal in each $\ell$-band (with arbitrary normalization).

The current data are in tension with even the high-feedback simulations for the CMB-derived best-fit cosmological parameter set.
Even with the kSZ effect set to zero, the tSZ scaling factor is only $0.55\pm0.21$ of what is predicted for the fiducial WMAP5 cosmology.  Meanwhile, the tSZ scaling factors are $0.42 \pm 0.21$ for the homogeneous kSZ model, and $0.34 \pm 0.21$ for the patchy kSZ model. The tension grows worse for the medium-feedback simulations by \citet{shaw09}.  Although the \citet{bode07} model is calibrated to reproduce observed x-ray scaling relations for high-mass, low-redshift clusters, it may significantly over-estimate the contribution of low-mass or high-redshift clusters (for which there are few direct x-ray observations with which to compare).

As mentioned above, an alternate interpretation of the low tSZ scaling factor is that the CMBall parameter chains without SZ constraints overestimate $\sigma_8$.
In order to constrain $\sigma_8$ based on the measurement of ${\rm A}_{\rm SZ}$, we need to construct the likelihood of observing ${\rm A}_{\rm SZ}$ for a given cosmological parameter set.
At a minimum, this likelihood would reflect the 12\% uncertainty due to sample variance of the tSZ effect. 
However given the large range of tSZ predictions, we add an additional theory uncertainty in quadrature with the sample variance.
The grey region in Figure~\ref{fig:theoryprior} shows the $1\,\sigma$ region
encompassed by a lognormal distribution of width $\sigma_{Asz} = 0.5$
around the fiducial model (black solid line). 
In the range where SPT is most sensitive ($2000 \le \ell \le 6000$), all the
predicted curves lie within this region. In order to account for the
range of predicted power spectra, we adopt a 50\%
theory uncertainty on the value of $\ln (A_{sz})$ in the likelihood
calculation. The resulting $\sigma_8$ constraints are dominated by the theory uncertainty. 
With this prior, we construct a new chain from the original parameter space through importance sampling.   
The regions preferred by this prior are shown in Figure \ref{fig:szdeplete}, as are the results of the new Markov chain. 

\begin{figure*}[ht]\centering
\includegraphics[width=0.9\textwidth]{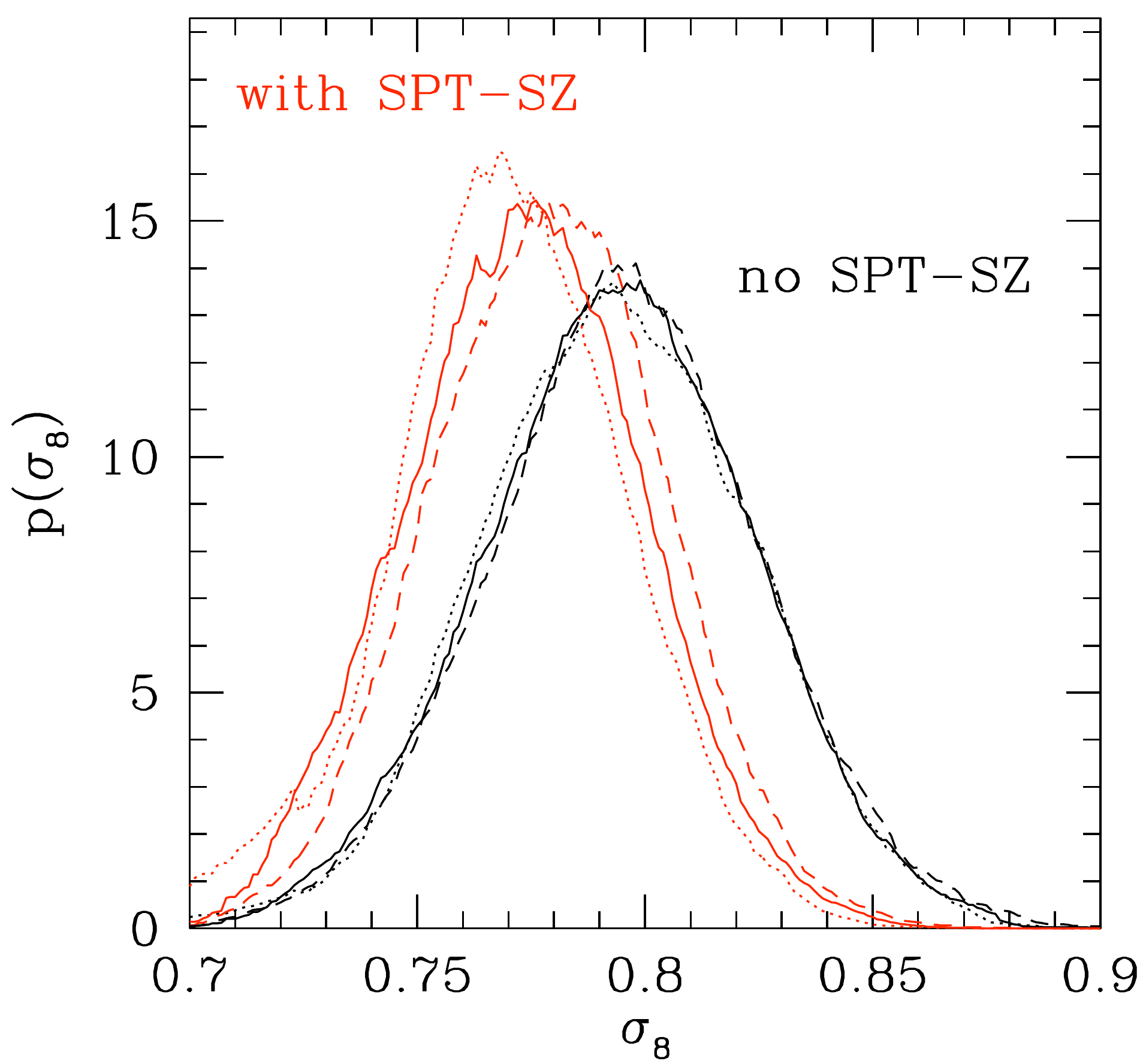}
  \caption[]{The 1D marginalized $\sigma_8$ constraints with and without including the SPT DSFG-subtracted bandpowers for three kSZ cases. 
  The {\bf black} lines denote the $\sigma_8$ constraints without SPT, while the {\bf red} lines include SPT's bandpowers.  
  Constraints with the patchy kSZ template are shown with a {\bf solid} line. 
  The results when including only the homogeneous kSZ model are shown with the {\bf dashed} lines, and the results for no kSZ effect are shown with {\bf dotted} lines. 
  The SPT data tightens the $\sigma_8$ constraint in all three cases. 
  }\label{fig:like1d_sigma8}
\end{figure*}

Constraints on $\sigma_8$ with and without including the SPT $A_{\rm SZ}$
measurements are shown in Table \ref{tab:sigma8} and Figure \ref{fig:like1d_sigma8}.
Under the assumption of the homogeneous kSZ model and a 50\% theoretical uncertainty
in the amplitude of the tSZ powerspectrum, the addition of the SPT data slightly
tightens the constraint on $\sigma_8$, while reducing the central value from
$\sigma_8=0.794 \pm 0.028$ to $0.773 \pm 0.025$. The uncertainty in the resulting
constraint on $\sigma_8$ is dominated by the large theoretical uncertainty in the
tSZ amplitude. If instead we assume that the fiducial tSZ model is perfectly
accurate and do not account for model uncertainty, the uncertainty on $\sigma_8$ is
reduced by 30\% and the preferred value is significantly reduced to $\sigma_8=0.746
\pm 0.017$ for the homogenous kSZ model. Despite the fact that the adopted template
is the lowest of the tSZ models shown in Figure~\ref{fig:szng}, the \citet{sehgal09} template, 
the value of $\sigma_8$ inferred by the SPT data using this model is lower than that
favored by WMAP. Additional SPT data will soon determine if the apparent tension
between $\sigma_8$ inferred from SZ and primordial CMB measurements is robust. In
any case, improving our theoretical understanding of both the kSZ and tSZ power
spectra is essential for fully realizing the potential of SZ power spectrum
measurements to constrain cosmological parameters such as $\sigma_8$.
 
\section{Conclusions}
\label{sec:conclusions}

We have presented the first CMB temperature anisotropy power spectrum results from the SPT experiment. 
Table \ref{tab:bandpowerssinglefreq} contains bandpowers based on simultaneous observations at 150 and $220\,$GHz of a single 100\,deg$^2$ field for three months during the austral winter of 2008. 
A total of 286k detector-hours of data was collected during these observations, resulting in a final map depth of 18\,$\mu$K-arcmin at 150$\,$GHz and 40\,$\mu$K-arcmin at 220$\,$GHz.  
The SPT 150$\,$GHz temperature maps are calibrated to an accuracy of 3.6\% through a direct comparison of CMB temperature anisotropy as observed by SPT and WMAP5. 
This absolute temperature calibration is extended to 220$\,$GHz with a 7.2\% uncertainty by an internal comparison of the primary CMB anisotropy measured by SPT at 150 and 220$\,$GHz.

The SPT bandpowers include angular multipoles from $\ell=2000$ to 9500. 
This broad angular range covers the transition from where the primary CMB anisotropy dominates the 
measured power to where secondary CMB anisotropies and foregrounds become dominant. 
The primary anisotropy observed by SPT is consistent with $\Lambda$CDM cosmological models based on lower-$\ell$ CMB data. 
At $\ell > 3000$, the SPT bandpowers are used to detect an excess of power above the primary anisotropies at $\sim$50$\,\sigma$ at both 150 and 220$\,$GHz after masking bright ($>6.4\,$mJy) sources.  

Most of this power can be explained by a population of faint dusty, star-forming galaxies (DSFGs), the properties of which are discussed in H09.  
We expect the DSFG population to contain a clustered component, which will produce a power term of comparable size to the tSZ effect at 150$\,$GHz. 
This clustered component would bias the measurement of power from secondary CMB anisotropies in single-frequency maps. 
However, we use SPT's multi-frequency data to construct the DSFG-subtracted bandpowers shown in Figure \ref{fig:cldif}. 
This frequency differencing reduces the clustered point source power to negligible levels and significantly reduces the Poisson power. 
  
Secondary CMB anisotropies, consisting of a linear combination of the kSZ and tSZ  effects, are detected at 2.6$\,\sigma$ in an analysis of the DSFG-subtracted bandpowers. 
Assuming there is no kSZ contribution, this equates to $4.2 \pm 1.5\,\mu$K$^2$ of tSZ power at $\ell = 3000$. 
This is substantially lower than the $8.5\,\mu$K$^2$ predicted by the homogeneous kSZ and fiducial tSZ models using WMAP5 cosmological parameters. 
The lower than expected value of the measured SZ power may point to either lower values of $\sigma_8$ or an over-estimate of the modeled SZ power spectrum.    
Adopting a conservative prior on the uncertainty in the model predictions for the SZ amplitude, we combine the measurement of the tSZ power spectrum with CMB data on larger angular scales to 
improve the constraint on the CMB-derived matter power spectrum normalization, $\sigma_8$. 
This results in a decrease in the preferred value to $\sigma_8 = 0.773 \pm 0.025$.  
A major source of uncertainty in this analysis is due to the current theoretical uncertainty on the amplitudes and shapes of the thermal and kinetic SZ effect power spectra.    
In the absence of this theoretical uncertainty, the constraint tightens about a still lower value of $\sigma_8 = 0.746\pm0.017$. 
The tension between the predicted and measured SZ amplitudes decreases for models that predict less kSZ or tSZ power, for instance models with shorter reionization epochs or higher energy feedback in clusters.
However, the tension remains at some level for all the models we considered.
It is our hope that the progress in observational efforts represented by this work will lead to further work
in improving theoretical models for secondary CMB anisotropies.  

The SPT power spectrum results presented here use only a small fraction of the complete data set.  
The SPT is continuing to take data and has already observed 800 deg$^2$ to similar depth at 
150 and 220$\,$GHz to the $100\,{\rm deg}^2$ used in this work, and 
600 deg$^2$ of this area also has $95\,$GHz coverage.  
The complete SPT survey is expected to cover over $2000\, {\rm deg}^2$ with these three frequencies. 
The complete data set promises a $>10\,\sigma$ detection of the tSZ power spectrum, and the addition 
of the third ($95\,$GHz) frequency band will allow us to separate the kSZ effect
from DSFGs and the tSZ effect.
Measuring the kSZ power spectrum will open a new window onto the reionization history of the universe.
The complete SPT data set will also produce a catalog of hundreds of SZ-selected clusters, high signal-to-noise
images of known clusters, a direct detection of gravitational lensing of the CMB, 
and the widest study to-date on the resolved and unresolved source populations at these frequencies.

\acknowledgments

The SPT team gratefully acknowledges the
contributions to the design and construction of the telescope by S.\ Busetti, E.\ Chauvin, 
T.\ Hughes, P.\ Huntley, and E.\ Nichols and his team of iron workers. We also thank the 
National Science Foundation (NSF) Office of Polar Programs, the United States Antarctic Program and the Raytheon Polar Services Company for their support of the project.  We are grateful for professional support from the staff of the South Pole station. We thank T. M.\ Lanting, J.\ Leong, A.\ Loehr, W.\ Lu, M.\ Runyan, D.\ Schwan, M.\ Sharp, and C.\ Greer for their early contributions to the SPT project,  J.\ Joseph and C.\ Vu for their contributions to the electronics and P. Ralph for his useful discussions and insights. 

The South Pole Telescope is supported by the National Science Foundation through grants ANT-0638937 and ANT-0130612.  Partial support is also provided by the  
NSF Physics Frontier Center grant PHY-0114422 to the Kavli Institute of Cosmological Physics at the University of Chicago, the Kavli Foundation and the Gordon and Betty Moore Foundation.
The McGill group acknowledges funding from the National Sciences and
Engineering Research Council of Canada, the Quebec Fonds de recherche
sur la nature et les technologies, and the Canadian Institute for
Advanced Research. The following individuals acknowledge additional support:
K.\ Schaffer and B.\ Benson from a KICP Fellowship, O.\ Zahn from a Berkeley Center for Cosmological Physics Fellowship, J.\ McMahon from a Fermi Fellowship,  Z.\ Staniszewski from a GAAN Fellowship, and A.T.\ Lee from the Miller Institute for Basic Research in Science,
University of California Berkeley.
N.W.\ Halverson acknowledges support from an Alfred P.\ Sloan Research
Fellowship.

This research used resources of the National Energy Research Scientific Computing Center, which is supported by the Office of Science of the U.S. Department of Energy under Contract No. DE-AC02-05CH11231. Some of the results in this paper have been derived using the HEALPix \citep{gorski05} package. We acknowledge the use of the Legacy Archive for Microwave Background Data Analysis (LAMBDA). Support for LAMBDA is provided by the NASA Office of Space Science.

\clearpage

\bibliography{../../BIBTEX/spt.bib}

\appendix

\section{Bandpower Covariance Matrix Estimation}\label{app:covmatrix} 

The bandpower covariance matrix includes both signal and noise contributions. 
The signal covariance is calculated from simulations. 
The noise covariance is estimated from the data. We calculate the variance of the mean power spectrum using the variance of cross-spectra between independent real maps.
With 300 independent observations of the same field, these maps are sufficient to generate an accurate estimate of the covariance matrix.     
Several details of our approach are motivated by the analytical treatments of \cite{tristram05} and \cite{polenta05}, and we first review the analytic estimate of the covariance matrix before discussing the estimator.

\subsection{Analytical Considerations}

Following \citet{tristram05}, we represent the expected covariance between two cross spectra as $\Xi$:  
\begin{align}
\Xi_{\ell\ell^\prime}^{AB,CD}&\equiv\left<\left(\widehat{D}_\ell^{AB}-\left<\widehat D_{\ell}^{AB}\right>\right)\left(\widehat D_{\ell^\prime}^{CD}-\left<\widehat D_{\ell^\prime}^{CD}\right>\right)\right>\\
&=\left(K[\textbf{W}]^{-1}\right)_{b^{\prime\prime}b}
\left(
\left<D^{AB}_{b^{\prime\prime}} D^{CD}_{b^{\prime\prime\prime}}\right>
-\left<D^{AB}_{b^{\prime\prime}}\right>\left<D^{CD}_{b^{\prime\prime\prime}}\right>
\right)
\left(K[\textbf{W}]^{-1}\right)_{b^{\prime\prime\prime}b^\prime}.
\end{align}
Our goal is to express this covariance in terms of the noise and signal in the maps, and to compute the magnitude of the diagonal elements, as well as the correlation between bandpowers. As the first step, the central term can be rewritten as,
\begin{align}
\left<D^{AB}_{b^{}} D^{CD}_{b^{\prime}}\right>
-\left<D^{AB}_{b^{}}\right>\left<D^{CD}_{b^{\prime}}\right>
&=\notag P_{bk}P_{b^\prime k^\prime}\frac{1}{(2\pi)^2}\int d\theta_k d\theta_{k^\prime}
\left(
\left<\widetilde{m}^{A}_{\textbf{k}^{}}\widetilde{m}^{B*}_{\textbf{k}^{}}
\widetilde{m}^{C}_{\textbf{k}^{\prime}}\widetilde{m}^{D*}_{\textbf{k}^{\prime}}\right>
-\left<\widetilde{m}^{A}_{\textbf{k}^{}}\widetilde{m}^{B*}_{\textbf{k}^{}}\right>
\left<\widetilde{m}^C_{\textbf{k}^{\prime}}\widetilde{m}^{D*}_{\textbf{k}^{\prime}}\right>
\right)\\
\label{eqn:dddd}
&=P_{bk}P_{b^\prime k^\prime}\frac{1}{(2\pi)^2}\int d\theta_k d\theta_{k^\prime}
\left(
\left<\widetilde{m}^{A}_{\textbf{k}^{}}\widetilde{m}^{C}_{\textbf{k}^{\prime}}\right>
\left<\widetilde{m}^{B*}_{\textbf{k}^{}}\widetilde{m}^{D*}_{\textbf{k}^{\prime}}\right>
+\left<\widetilde{m}^{A}_{\textbf{k}^{}}\widetilde{m}^{D*}_{\textbf{k}^{\prime}}\right>
\left<\widetilde{m}^{B*}_{\textbf{k}^{}}\widetilde{m}^{C}_{\textbf{k}^{\prime}}\right>
\right).
\end{align}

To simplify equation \ref{eqn:dddd}, \cite{tristram05} and \cite{polenta05}  make the following assumptions:
\begin{enumerate}
\item Fluctuations in the map, i.e., from CMB anisotropies, confusion limited point sources and noise, are well described by a Gaussian random field. 
\item The beams and filtering applied to the data are isotropic; $E_\textbf{k} \equiv G_\textbf{k}B_\textbf{k}$ depends only on $|\textbf{k}|$.
\item The instrumental noise is isotropic.
\item The power spectrum, $C_{k}$, is smoothly varying with $k$, and changes little over scales comparable to the width of the mode coupling matrix. 
\end{enumerate}
Using the assumption that $C_k$ and $|E_\textbf{k}|$ do not vary much over small changes in $\textbf{k}$, these products become:
\begin{align}
\label{eqn:mmprod}
\left< \widetilde{m}^{A}_\textbf{k}\widetilde{m}^{B*}_{\textbf{k}^\prime}\right>
=\sum_{\textbf{k}^{\prime\prime}\textbf{k}^{\prime\prime\prime}}
\widetilde{\textbf{W}}_{\textbf{k}^{}-\textbf{k}^{\prime\prime}}
\widetilde{\textbf{W}}^{*}_{\textbf{k}^{\prime}-\textbf{k}^{\prime\prime\prime}}
E_{\textbf{k}^{\prime\prime}}E^{*}_{\textbf{k}^{\prime\prime\prime}}
\left<a^{A}_{\textbf{k}^{\prime\prime}}a^{B*}_{\textbf{k}^{\prime\prime\prime}}\right>
=\sum_{\textbf{k}^{\prime\prime}}
\widetilde{\textbf{W}}_{\textbf{k}^{}-\textbf{k}^{\prime\prime}}
\widetilde{\textbf{W}}_{\textbf{k}^{\prime\prime}-\textbf{k}^\prime}
\left|E_{\textbf{k}^{\prime\prime}}\right|^2 C^{AB}_{k^{\prime\prime}}
\approx \widetilde{\textbf{W}^2}_{\textbf{k}^{}-\textbf{k}^\prime}\left|E_{\textbf{k}}\right|^2C^{AB}_{k}
\end{align}
Here $C^{AB}_k$ is shorthand for $C_k+\frac{N^{A}_k}{B_k^2}\delta_{AB}$, the expected cross spectrum between two unfiltered, perfectly beam-corrected---though noisy---maps.  
The additional term is the noise bias that exists in the map auto-spectrum.  
Assuming isotropic beams and filtering, we can combine equations \ref{eqn:dddd} and \ref{eqn:mmprod} to obtain a relatively simple expression:
\begin{align}
\left<D^{AB}_{b^{}} D^{CD}_{b^{\prime}}\right>
-\left<D^{AB}_{b^{}}\right>\left<D^{CD}_{b^{\prime}}\right>
&=P_{bk}P_{b^\prime k^\prime}\frac{1}{\left(2\pi\right)^2}\int d\theta_k d\theta_{k^\prime} 
\left|\widetilde{\textbf{W}^2}\right|^2\left|E_{k}\right|^2\left|E_{k^\prime}\right|^2
\left(C_k^{AB}C_{k^\prime}^{CD}+C_k^{AD}C_{k^\prime}^{BC}\right) \nonumber \\
\label{eqn:covrawanalytic}&=P_{bk}P_{b^\prime k^\prime}E_k^2 E_{k^\prime}^2 M[\textbf{W}^2]_{kk^\prime}\left(C_k^{AC}C_{k^\prime}^{BD}+C^{AD}_{k}C^{BC}_{k^\prime}\right).
\end{align}

Combining equations \ref{eqn:dddd} and \ref{eqn:covrawanalytic} yields:

\begin{align}
\Xi^{AB,CD}_{bb^\prime}=\left(K[\textbf{W}]^{-1}\right)_{b^{(2)}b}\left[P_{b^{(2)}k}P_{b^{(3)} k^\prime}M[\textbf{W}^2]_{kk^\prime}E_{k}^2E_{k^\prime}^2\left(C_k^{AC}C_{k^\prime}^{BD}+C^{AD}_{k}C^{BC}_{k^\prime}\right)\right]\left(K[\textbf{W}]^{-1}\right)_{b^{(3)}b^\prime}.
\end{align}
To obtain a simplified expression for the magnitude of the diagonal elements of the covariance matrix, one typically assumes the mode-coupling matrix is nearly diagonal: $M_{kk^\prime}[\textbf{W}]\approx w_2 \delta_{kk^\prime}$. 
 In this approximation $\left(K[\textbf{W}]^{-1}\right)_{bb^\prime}\approx w_2^{-1}E_b^{-2}\delta_{bb^\prime}$ and $M_{kk^\prime}[\textbf{W}^2]\approx w_4 \delta_{kk^\prime}$ and thus the covariance of any two cross spectra is:
\begin{align}
\Xi^{AB, CD}_{bb^\prime}&\approx\frac{w_4}{N_b w_2^2}\left(\frac{\ell_{\textrm{eff},b}(\ell_{\textrm{eff},b}+1)}{2\pi}\right)^2\left(C^{AC}_bC^{BD}_{b}+C^{AD}_bC^{BC}_b\right)\delta_{bb^\prime}\nonumber\\
\label{eqn:XiGeneral}&=\frac{\left<\widehat{D}^{AC}_b\right>\left<\widehat{D}^{BD}_{b}\right>+\left<\widehat{D}^{AD}_b\right>\left<\widehat{D}^{BC}_b\right>}{\nu_b}\delta_{bb^\prime},
\end{align}
where $\nu_b$ is the effective number of independent k-modes in each $\ell$-band.  For isotropic filtering, $\nu^{\textrm{iso}}_b=\frac{N_b w_2^2}{w_4}$.

One subtlety in estimating the covariance is the fact that although the noise in each map is independent, each map has the same sky coverage. Hence the signal in all maps is correlated.  
The correct estimator must take this correlation into account. 
Under the simplifying assumption that all maps are statistically equivalent,  the correlation between two cross spectra depends only on whether the spectra have a map in common (e.g. the cross spectrum $D_b^{12}$ is more strongly correlated to the spectrum $D_b^{23}$ than $D_b^{34}$).   
Comparing the covariance of two cross-spectra taken from 4 different observations:
\begin{align}
\label{eqn:allunique}
\Xi_{bb}^{AB, CD}|_{A\ne B\ne C\ne D}&=\frac{2}{\nu_b}C^2_b,
\end{align}
to the covariance of a pair of cross-spectra with a common map,
\begin{align}
\label{eqn:oneunique}
\Xi_{bb}^{AB, BC}|_{A\ne B\ne C}\
&=\frac{1}{\nu_b}\left(2C_b^2+\frac{N_b}{B_b^2}C_b\right),
\end{align}
to the covariance of a pair of cross-spectra with two maps in common,
\begin{align}
\label{eqn:noneunique}
\Xi_{bb}^{AB, AB}|_{A\ne B}
&=\frac{1}{\nu_b}\left(2C_b^2+2\frac{N_b}{B_b^2}C_b+\frac{N^2_b}{B^4_b}\right).
\end{align}
The degree of correlation is also $\ell$ dependent since it depends on the relative signal vs. noise power in the maps.  
In the high-$\ell$ regime, where noise dominates the power in an individual map, all cross spectra are nearly independent. 
Conversely all cross-spectra are nearly completely correlated at low-$\ell$, where the primary CMB anisotropy overwhelms the noise. 

Given the assumption of statistical equivalence, we can then compute the expected variance of the mean spectrum based on these variance estimates. 
This is equal to the correlation between any particular cross-spectrum and the mean,
\begin{equation}
\label{eqn:meanvariance}
\Xi_{bb}^{\textrm{mean},\textrm{mean}}=\Xi_{bb}^{\textrm{mean}, AB}|_{A\ne B}\approx\frac{1}{\nu_b}\left(2C_b^2+4C_b \frac{N_b}{n_{obs} B_b^2}+2\frac{N_b^2}{n_{obs}^2 B_b^4}\right).
\end{equation}
 This estimate of the variation of the mean spectrum agrees with the uncertainty estimates given in \cite{polenta05}.

It should be noted that the noise and filtering of the SPT data are anisotropic.  
Equation \ref{eqn:XiGeneral} can therefore be used only as a guideline.
In the case of anisotropic filtering or anisotropic beams, the number of independent modes per bin will be typically smaller than $\nu^{\textrm{iso}}_b$, since anisotropic filtering will weight different k-space modes unevenly.   
For example, the k-space mask completely eliminates all modes with $k_x<1200$.  
The variance in each $\ell$-bin increases with fewer independent modes.
However, even if we account for the effective number of independent modes in an $\ell$-bin, equation \ref{eqn:XiGeneral} does not account for the anisotropic nature of the atmospheric noise contribution.

\subsection{The Empirical Covariance Estimator}

The existing analytic treatments are not directly applicable to the SPT data due to anisotropies in the noise and filters. 
By the nature of SPT's scan strategy, atmospheric fluctuations preferentially contaminate low $k_x$ modes.  
Likewise the filters intended to remove these fluctuations preferentially remove low $k_x$ modes.  
Instead, we have designed an empirical estimator which reproduces the analytical results when applied to isotropic data, while accurately accounting for the increased uncertainty due to the noise and filtering anisotropies in the actual data. 

The noise covariance matrix estimate is divided into two parts, a signal contribution obtained from the Monte Carlo simulations described in \S\ref{sec:simulations} and a noise contribution obtained from real single-observation maps:
\begin{equation}
\label{eqn:covdeftop}
\textbf{C}^{}_{bb^\prime}=\textbf{C}^{\textrm{MC,s}}_{bb^\prime}+\textbf{C}^{\textrm{data}}_{bb^\prime}.
\end{equation}
The signal contribution is straightforward to estimate with an approach similar to the MASTER power spectrum error estimator. We use the signal only simulations to obtain an empirical estimate of the sample variance:
\begin{equation}
\label{eqn:covsig}
\textbf{C}^\textrm{MC,s}_{bb^\prime}=\overline{\Delta \widehat{D}^\textrm{MC, s}_{b}\Delta \widehat{D}^\textrm{MC, s}_{b^{\prime}} }.
\end{equation} 
Note that here $\Delta x\equiv x - \overline{x}$ is defined with respect to the sample mean.  
Since the simulations include only CMB realizations and point sources in the confusion limit, the simulated signals are essentially Gaussian. Therefore we expect the usual sample variance contribution:
\begin{equation}
\label{eqn:expsamplevar}
\left<\textbf{C}^{}_{bb^\prime, (\textrm{MC,s})}\right>=\frac{2 C_b^\textrm{theory}}{\nu_b}.
\end{equation}
As before, $\nu_b$ is the effective number of independent Fourier-modes in each $\ell$-band.   

The noise contribution is computed from the cross spectra of single-observation maps. 
We use the following estimator for the noise contribution:

\begin{equation}
\label{eqn:datacovdef}
\textbf{C}_{bb^\prime}^\textrm{data}\equiv \frac{2f(n_\textrm{obs})}{n^4_\textrm{obs}}\sum_{\lambda}\sum_{\alpha\ne\lambda}
\left(
\Delta \widehat{D}_{b}^{\lambda\alpha}\Delta \widehat{D}_{b^\prime}^{\lambda\alpha}+
2\left[\sum_{\beta\ne\lambda, \alpha}\Delta \widehat{D}_{b}^{\lambda\alpha}\Delta \widehat{D}_{b^\prime}^{\lambda\beta}
\right]
\right).
\end{equation}
Here $f(n_\textrm{obs})$ is a correction due to the finite number of realizations.  
In the limit of many observations this function asymptotes to unity;  we use 300 observations so this term can be ignored.  
The first term can be identified as the sample variance of the cross spectra.  
The second term accounts for the additional correlations between cross-spectra with a common map.

We can now calculate the expectation value for the noise component of the covariance estimator defined in \ref{eqn:datacovdef}:

\begin{equation}
\label{eqn:expnoise}
\left<\textbf{C}^\textrm{data}_{bb}\right>\approx\frac{2}{n_{obs}^2}\Xi_{bb}^{\lambda\alpha,\lambda\alpha}+\frac{4}{n_{obs}}\Xi_{bb}^{\lambda\alpha,\lambda\beta}-\left(\frac{4}{n_{obs}}+\frac{2}{n_{obs}^2}\right)\Xi^{\textrm{mean},\textrm{mean}}=\frac{1}{\nu_b}\left(4C_b\frac{N_b}{n_{obs} B_b^2}+2\frac{N_b^2}{n_{obs}^2B_b^4}\right).
\end{equation}

This is combined with the signal, or cosmic variance, component in equation \ref{eqn:covsig} to get the
expectation value of the estimator:
\begin{equation}
\left<\textbf{C}_{bb}\right>\approx\frac{1}{\nu_b}\left[2 {C_b^\textrm{theory}}^2+4C_b\frac{N_b}{n_{obs} B_b^2}+2\frac{N_b^2}{n_{obs}^2B_b^4}\right].
\end{equation}
This agrees with the analytic estimate (equation \ref{eqn:meanvariance}) for the variance of the mean.

\subsection{Multifrequency Cross Covariances}
A multifrequency data set requires an estimate of both the covariance of the each individual set of bandpowers (i.e., both the single-frequency bandpowers, and the cross-frequency bandpowers) and the cross-covariance between these sets.  
We naturally expect signal correlations between different sets of bandpower due to the fact that all three sets of bandpowers reported here are derived from the same patch of sky.   
However we also expect \emph{noise} correlations between the $\textrm{150$\,$GHz}\times\textrm{220$\,$GHz}$ cross spectrum bandpowers and each set of single-frequency bandpowers since the noise uncertainty in the cross-spectrum is entirely due to noise in the 150$\,$GHz and 220$\,$GHz data.   
Thus we compute the cross-covariance matrices, ${\textbf{C}_{bb^\prime}}^{(i,j)}$, where i and j denote one of the three sets of bandpowers: 150$\,$GHz, 220$\,$GHz and 150$\,$GHz$\times$220$\,$GHz: 
\begin{equation}
\label{eqn:covdefmulti}
{\textbf{C}_{bb^\prime}}^{(i,j)}=\overline{{\Delta \widehat{D}^\textrm{MC, s}_{b}}^{(i)}{\Delta \widehat{D}^\textrm{MC, s}_{b^{\prime}}}^{(j)} }+\frac{2f(n_{obs})}{n^4_m}\sum_{\lambda}\sum_{\alpha\ne\lambda}
\left(
{\Delta \widehat{D}_{b}^{\lambda\alpha}}^{(i)}{\Delta \widehat{D}_{b^\prime}^{\lambda\alpha}}^{(j)}+
2\left[\sum_{\beta\ne\lambda, \alpha}{\Delta \widehat{D}_{b}^{\lambda\alpha}}^{(i)}{\Delta \widehat{D}_{b^\prime}^{\lambda\beta}}^{(j)}
\right]
\right).
\end{equation}

\subsection{Treatment of Off-diagonal Elements}
  Given the finite number of simulations and data maps, we expect some statistical uncertainty in the covariance estimate, particularly in the off-diagonal elements.  
  Such uncertainty is not unique to the estimation technique described here, rather it is expected for \emph{any} covariance estimate which is computed from a finite number of realizations.  
  We expect the covariance estimates to be Wishart distributed, with $n_{obs}=300$ degrees of freedom.    
   A given covariance element, $\textbf{C}_{ij}$ has a statistical variance of:
\begin{equation}
\left<\left(\textbf{C}_{ij}-\left<\textbf{C}_{ij}\right>\right)^2\right>=\frac{\textbf{C}_{ij}^2+\textbf{C}_{ii}\textbf{C}_{jj}}{n_{obs}}.
\end{equation}
For diagonal elements we expect a standard deviation of $\sqrt{2/n_{obs}}=1/\sqrt{150}=8.1\%$.  In addition, there is a statistical uncertainty on the apparent correlation between two bins.  
If we assume that the true correlation between bins is small, then the standard deviation of the apparent correlation between two bins is $\sqrt{1/n_{obs}}$ or 5.7\%.  
For the choice of bin-size, the statistical error on the correlation of any two bins is much larger than the expected correlation (i.e. the fractional error on the apparent correlation estimates is greater than 100\% even for adjacent bins).  For bins that are widely separated, these false bin-bin correlations may skew model fitting.   
Therefore we ``condition" the published covariance matrices in order to reduce this statistical uncertainty on the covariance matrices.
 
From equation \ref{eqn:covrawanalytic}, we see that the shape of the correlation matrix (i.e. the relative size of the on-diagonal to off-diagonal covariance elements as a function of bin separation) is determined by the apodization window through the quadratic mode-coupling matrix, $M[\textbf{W}]_{kk}$.   
For the $\ell$ range considered in this work, the off-diagonal elements of this matrix depend only on the distance from the diagonal, $|k-k^\prime|$.   
Therefore we condition the estimated covariance matrix, $\widehat{\textbf{C}}_{kk^\prime}$ by first computing the corresponding correlation matrix, and then averaging all off-diagonal elements of a fixed separation from the diagonal:
\begin{equation}
\label{eqn:covcond}
\textbf{C}^\prime_{kk^\prime}=\frac{
\sum_{k_1-k_2=k-k^\prime} 
\frac{\widehat{\textbf{C}}_{k_1k_2}}{\sqrt{\widehat{\textbf{C}}_{k_1k_1}\widehat{\textbf{C}}_{k_2k_2}}}
}{\sum_{k_1-k_2=k-k^\prime} 1}.
\end{equation}
The bandpowers reported in Tables \ref{tab:bandpowerssinglefreq} and \ref{tab:bandpowers} are obtained by first computing power spectra and covariance matrices for a bin-width of $\Delta\ell=100$ with a total of 80 preliminary bins.  
This covariance matrix was then conditioned according to equation \ref{eqn:covcond} before averaging the bandpowers and covariance matrix into the final bands.
	
Equation \ref{eqn:covrawanalytic} is based on the assumption that the filtering is isotropic. In order to test the validity of this equation for the anisotropic filtering, we perform 10000 simple Monte Carlo simulations.  
In each simulation a white noise realization is subjected to a simplified, though similarly anisotropic, version of the filtering scheme.   
The variance of the resultant spectra is computed and compared to equation \ref{eqn:covrawanalytic}.   
Though the apparent correlations between all bins exhibit the expected 1\% scatter, the correlations between neighboring bins are consistent with equation \ref{eqn:covrawanalytic}.

\end{document}